\documentclass[useAMS]{mn2e}

\usepackage{times}
\usepackage{epsfig,amssymb}
\usepackage{graphics}
\usepackage{graphicx}
\usepackage{amsfonts}
\usepackage{eufrak}
\usepackage{mathrsfs}
\usepackage{eucal}
\usepackage{color}
\usepackage{makeidx}
\usepackage{showidx}
\usepackage{amsmath}
\usepackage{fontenc}
\def\AaA{{AA}}
\def\aj{{AJ}}
\def\apj{{ApJ}}
\def\apjl{{ApJ}}
\def\apjs{{Astrophys.~J.~Suppl.}}
\def\araa{{ARA\&A}}
\def\mnras{{MNRAS}} 
\def\nat{{Nat}}

\def\prd{{PhRvD}}
\def\etal{et al.~}  

\newcommand{\vta}[1]{\vert \boldsymbol{a}_{#1}\vert}
\newcommand{\vtl}{\vert \boldsymbol{{\ell}}\vert}
\newcommand\ergsec{\ifmmode \mathrm{erg~s}^{-1} \else erg~s$^{-1}$\fi}
\newcommand\Msun{\ifmmode \mathrm M_{\odot} \else $M_{\odot}$\fi}
\newcommand\MBH{\ifmmode M_{\mathrm{bh}} \else $M_{\mathrm{bh}}$\fi}
\newcommand\JBH{\ifmmode J_{\mathrm{bh}} \else $J_{\mathrm{bh}}$\fi}
\newcommand\JD{\ifmmode J_{\mathrm{d}} \else $J_{\mathrm{d}}$\fi}
\newcommand\Jt{\ifmmode J_{\mathrm{tot}} \else $J_{\mathrm{tot}}$\fi}
\newcommand\Rw{\ifmmode R_{\mathrm{warp}} \else $R_{\mathrm{warp}}$\fi}
\newcommand\Rsch{\ifmmode R_{\mathrm{Schw}} \else $R_{\mathrm{Schw}}$\fi}
\newcommand\Rsg{\ifmmode R_{\mathrm{sg}} \else $R_{\mathrm{sg}}$\fi}
\newcommand\vjh{\boldsymbol{J}_{\mathrm{bh}}}
\newcommand\vjd{\boldsymbol{J}_{\mathrm{d}}}
\newcommand\vjt{\boldsymbol{J}_{\mathrm{tot}}}
\newcommand\Macc{\ifmmode M_{\mathrm{acc}} \else $M_{\mathrm{acc}}$\fi}
\newcommand\Msg{\ifmmode M_{\mathrm{sg}} \else $M_{\mathrm{sg}}$\fi}
\newcommand\Md{\ifmmode M_{\mathrm{d}} \else $M_{\mathrm{d}}$\fi}
\newcommand\af{\ifmmode a^{\mathrm{f}} \else $a^{\mathrm{f}}$\fi}
\newcommand\te{\ifmmode \tilde{e} \else $\tilde{e}$\fi}
\newcommand\tl{\ifmmode \tilde{l} \else $\tilde{l}$\fi}
\newcommand{\rlso}{r_{\mathrm{lso}}}
\newcommand{\rdl}{\mathcal R}
\newcommand\Rg{\ifmmode R_{\mathrm{g}} \else $R_{\mathrm{g}}$\fi}
\newcommand{\hrlso}{\hat{r}_{\mathrm{lso}}}

\title[Grand unification of AGN activity in the $\Lambda$CDM cosmology]{Grand unification of AGN activity in the $\Lambda$CDM cosmology}
\author[Fanidakis \etal]{N.  Fanidakis,$^{1}$\thanks{E-mail: nikolaos.fanidakis@dur.ac.uk}
C. M. Baugh,$^{1}$ A. J. Benson,$^{2}$ R. G. Bower,$^{1}$ S. Cole,$^{1}$
\newauthor C. Done,$^{1}$ C. S. Frenk$^{1}$\\
$^{1}$Institute for Computational Cosmology, Department of Physics, University of Durham,\\
Science Laboratories, South Road, Durham DH1 3LE, United Kingdom\\
$^{2}$Mail Code 130-33, California Institute of Technology, Pasadena, CA 91125, USA}
\begin{document}    

\maketitle

\label{firstpage}

\begin{abstract}
We track the coevolution of supermassive black holes (SMBHs) and
their host galaxies through cosmic time. The calculation is embedded
in the \texttt{GALFORM} semi-analytic model which simulates the
formation and evolution of galaxies in a cold dark matter (CDM)
universe. The black hole (BH) and galaxy formation models are
coupled: during the evolution of the host galaxy, hot and cold gas
are added to the SMBH by flows triggered by halo gas cooling, disc
instabilities and galaxy mergers. This builds up the mass and spin
of the BH, and the resulting accretion power regulates gas
cooling and subsequent star formation. The accretion flow is assumed
to form a geometrically thin cool disc when the accretion rate
exceeds $0.01\dot{M}_{\mathrm{Edd}}$, and a geometrically thick,
radiatively inefficient hot flow when the accretion rate falls below
this value. The resulting quasar optical luminosity function matches
observations well, and the mass of the SMBH correlates with the
mass of the galaxy bulge as in the observed
$\MBH-M_{\mathrm{bulge}}$ relation. The BH spin
distribution depends strongly on whether we assume that the gas in any given
accretion episode remains in the same plane or it fragments
into multiple, randomly aligned accretion episodes due to its
self-gravity. We refer to these cases as the ``prolonged" and
``chaotic" accretion modes respectively. In the chaotic accretion
model there is a clear correlation of spin with SMBH mass (and hence
host galaxy bulge mass). Massive BHs ($M>5\times10^8~\Msun$) are
hosted by giant elliptical galaxies and are rapidly spinning, while
lower mass BHs are hosted in spiral galaxies and have much lower
spin. Using the Blandford--Znajek mechanism for jet production to
calculate the jet power, our model reproduces the radio
loudness of radio galaxies, LINERS and Seyferts, suggesting that the
jet properties of active galaxy nuclei (AGN) are a natural
consequence of {\em both} the accretion rate onto {\em and} the spin
of the central SMBH. This is the first confirmation that a CDM
galaxy formation model can reproduce the observed radio phenomenology of
AGN.
\end{abstract}

\begin{keywords}
 galaxies:jets -- galaxies:nuclei -- galaxies:active -- quasars:general -- methods:numerical
\end{keywords}

\section{Introduction} \label{sec:introduction}

Active galaxies can be classified according to the importance of the
radio emission from their nucleus. Objects showing strong emission
at radio wavelengths belong to the ``radio-loud" class, whereas
those with negligible emission belong to the ``radio-quiet" class.
Radio-loud AGN are associated with large-scale radio-emitting jets
while radio-quiet AGN show very little or negligible jet
activity. Sikora \etal (2007) found that AGN have a bimodal
distribution on the radio-optical plane, with radio-loud objects
being about $10^{3}$ times brighter in radio than radio-quiet
objects (see also Kellermann \etal 1989; Xu
\etal 1999). The origin of this dichotomy remains unknown. However,
it has been proposed in many studies that a first step towards
explaining why some AGN launch prominent jets while others do not is
to understand the nature of the central engine, the accreting BH.

AGN jets are believed to extract rotational energy from the BH and
accretion disc through magnetic fields. Analytical studies have
contributed to an understanding of the nature of the jets (Blandford
\& Znajek 1977; Macdonald \& Thorne 1982; Blandford \&
Payne 1982; Begelman, Blandford \& Rees 1984), but a breakthrough
has come from magneto-hydrodynamical (MHD) simulations of the
accretion flow. These self-consistently calculate the turbulent
magnetic field dynamo which is the physical origin of the stresses
which transport angular momentum outwards, allowing material to
accrete onto the BH (Balbus \& Hawley 1998). Embedding such
calculations in a fully general relativistic framework produces
relativistic jets from the accretion flow, and the collimation and
jet power depend on the BH spin (McKinney \& Gammie 2004; De
Villiers \etal 2005; Hawley \& Krolik 2006).

Thus, it now seems clear that the jet power depends on the vertical
(poloidal) magnetic field strength close to the BH, but the
simulations are still not yet at the level where they can predict
this {\it ab. initio} (e.g. Beckwith \etal 2008), so analytic models
are still required. The most popular of these, the
Blandford--Znajek (hereafter BZ) mechanism, uses the magnetic field
as a means to tap the spin of the BH. This gives a strong spin
dependence on the jet power.  Indeed, many authors have proposed
spin as the physical parameter that determines the radio loudness of
an AGN (see Wilson \& Colbert 1995; Hughes \& Blandford 2003). The
spin paradigm, as this idea is called, offers a plausible theoretical
explanation for the wide range of jet luminosities in AGN and could
be the basis for understanding the observed dichotomy between
radio-loud and radio-quiet AGN.

Alternatively, the radio--loud, radio--quiet switch may just be
determined by the physical state of the accretion flow. Stellar--mass BH binary
systems in our galaxy show a clear spectral transition at about 1\%
of the Eddington luminosity, $L\sim 0.01L_{\mathrm{Edd}}$, from a
hot, optically thin accretion flow at low luminosities (e.g. an
advection-dominated accretion flow or ADAF; Narayan \& Yi 1994) to a
cool, geometrically thin disc (Shakura \& Sunyaev 1973) at higher
luminosities (Esin \etal 1997, see e.g. the review by Done,
Gierlinski \& Kubota 2007). The collapse of the radio jet observed
across this transition clearly relates to the large drop in pressure
(and hence scale height, $H$) between the hot and cool flow (e.g.
 Fender, Belloni \& Gallo 2004). Even without BH spin, this produces a clear
dichotomy of radio properties (e.g. Jester 2005).

It seems most plausible that {\em both} these mechanisms, along with
the mass accretion rate, affect jet power.  The BZ mechanism has a
``hidden" dependence on mass accretion rate and the scale height of
the flow because the magnetic field strength close to the BH
saturates, due to the dynamo, into rough equipartition with the
pressure in the flow. This depends on the mass accretion rate for
either the disc or the hot flow, but the pressure in the hot flow is
much larger than that in the cool disc so the field strength and
hence jet power abruptly drop at this transition (Meier 2001; 2002)

Sikora \etal (2007) used these ideas to interpret observations of
radio galaxies and concluded that the data could be explained if the
jet depends on mass accretion rate, accretion model and
spin. However, they also required a mass-spin correlation, such that
the most massive BHs, which, according to the
$\MBH-M_{\mathrm{bulge}}$ relationship (Magorrian \etal 1998; 
McLure \& Dunlop 2002; Marconi \& Hunt 2003; 
Haring \& Rix 2004), reside in massive elliptical
galaxies, have higher spin than lower mass BHs, which reside in
spirals. They speculated that this could occur through the
hierarchical growth of structure, where the major mergers which give
rise to giant elliptical galaxies trigger large amounts of gas
accretion with a given angular momentum direction onto the BH,
spinning it up to the maximal value. By contrast, spiral galaxies
have not experienced a recent major merger, and grow mainly through
smooth accretion and multiple minor mergers with random angular momentum, resulting in
low spin and hence weak jet luminosities. 

Some of these ideas have recently been explored using simplified
models of galaxy formation by Berti \& Volonteri (2008) and Lagos
\etal (2009). In this paper, we calculate 
the growth of BHs by accretion and mergers, their acquisition of
spin and their accretion rates within a specific theory of galaxy
formation in a CDM model which has been extensively tested against a
large range of observations of the galaxy population. Not only is
our calculation of the joint evolution of BHs and galaxies
self-consistent, but we also incorporate several accretion and
jet-launching models from the literature which allows us to perform
a quantitative comparison with observations of AGN.

The paper is organised as follows. In Section~2, we explain how we
calculate the growth of the mass and spin of SMBHs through the
accretion of hot gas from the halo (radio mode) and cold gas from
the galactic disc and mergers (quasar mode). Most BHs grow primarily
by gas accretion except the most massive ones which form late and
increase their mass substantially by merging with other SMBHs. A BH
is spun up to approximately the maximal value when it accretes its
own mass from a flow at constant angular momentum. 

In Section~3, we present the resulting distributions of BH mass, gas
accretion rate and spin. We consider two possible modes of
accretion.  In the ``prolonged accretion" case, the gas is accreted
in a single episode. Even minor mergers trigger gas flows onto the
nucleus that often deposit a mass greater than the mass of the recipient
BH. Thus, most BHs are typically spun up to the maximal value. This
model predicts radio properties that do not reproduce the observed luminosity function. We then consider a ``chaotic accretion'' case in which
the accretion episodes are limited by the self-gravity of the disc. Gas flows onto the nucleus give rise to a series of accretion episodes each typically augmenting the BH mass by only a small factor. Successive
accretion events are uncorrelated, resulting in low spins for the
relatively low mass BHs which grow primarily by accretion. On the
other hand, the most massive holes which build up substantial mass
through BH-BH mergers are spun up to high (but not maximal) spin
values. In Section~4, we show that adopting this accretion model,
the optical luminosity from disc-accreting objects matches the
quasar luminosity function reasonably well.

In Section~5, we incorporate an explicit BZ model for the jet power
(Meier 2002) and use this to predict the radio luminosity function
which, in the chaotic accretion case, agrees well with
observations. On the basis of our derived optical and radio
luminosity distributions, we propose, in Section~6, a ``grand
unification of AGN activity", in which the accretion flow and jet
luminosity are related to AGN type through their mass, spin and mass
accretion rate. Fundamentally, our calculation shows, for the first
time, how the coeval growth of BHs and their host galaxies result in
optical and radio properties that explain the AGN
activity seen in the local Universe.

\section{Evolution of spin in hierarchical galaxy formation models}
\label{sec:Cosmological model}  

In this Section we describe the cosmological processes that initiate
and regulate the growth of SMBHs and outline the modelling of the
spin evolution due to gas accretion and mergers in hierarchical
cosmologies. Our basic modelling tool for our predictions is the
\texttt{GALFORM}
\normalsize code (Cole \etal 2000), and we use an update of
the Bower \etal (2006) version for modelling the formation and
evolution of galaxies in the $\Lambda$CDM cosmology. The changes
relative to the Bower \etal model are the following. Firstly,
the fraction, $\epsilon_{\mathrm{SMBH}}$, of the Eddington
luminosity of an accreting SMBH that is available for heating the
halo during an episode of AGN feedback is set to 0.01 (Bower \etal 2006 use,
$\epsilon_{\mathrm{SMBH}}=0.04$\footnote{Note that due to an error in Bower
\etal (2006), cooling luminosities were overestimated by a factor of
4$\pi$. Thus, while the paper quotes the efficiency parameter
$\epsilon_{\mathrm{SMBH}}$ as 0.5, this should have been $0.5/4\pi =
0.04$. With this correction, the rest of the parameters and results
are unchanged.}). Secondly, in starbursts triggered by galaxy mergers
or disc instabilities, we assume that the fraction,
$F_{\mathrm{SMBH}}$, of the cold gas turned into stars that
is accreted onto the BH is 0.01 (Bower \etal 2006 use
$F_{\mathrm{SMBH}}=0.017$). These changes are introduced to improve 
the modelling in this paper and \emph{do not} affect the fundamental
predictions of the Bower \etal model.

BHs are assumed to evolve in mass in accordance with the model developed in Malbon
\etal (2007) and Bower \etal (2006). We study the cosmological spin
evolution of SMBH seeds of a total of $\sim4.2\times10^{6}$
galaxies identified in the Millennium N-body simulation (co-moving volume of
$1.25\times10^{8}~h^{-3}$Mpc$^{3}$) from redshift 127 to redshift zero
(Springel \etal 2005). Here $h$ is defined by
$H_{0}=h\times100$~km~s$^{-1}$Mpc$^{-1}$, where $H_{0}$ is the Hubble
constant at redshift zero. The cosmology adopted in the simulations is
$h=0.72$, $\Omega_{\mathrm{m}}=0.25$, $\Omega_{\mathrm{b}}=0.045$,
$\Omega_{\mathrm{\Lambda}}=0.75$ and
$\sigma_8=0.9$\footnote{$\Omega_{\mathrm{m}}$, $\Omega_{\mathrm{b}}$
and $\Omega_{\mathrm{\Lambda}}$ express the present density of the
baryonic, total matter and dark energy components of the Universe
relative to the critical density ($\rho_{\mathrm{crit}}=3H^2/8\pi
G$). $\sigma_8$ measures the rms mass fluctuations in spheres of
radius $8~h^{-1}\mathrm{Mpc}$ linearly extrapolated to the present
day.}.

\subsection{The growth of SMBHs}

The evolution of BH mass fits naturally in the CDM
model of galaxy formation (Kauffmann \& Haehnelt 2000; Malbon \etal 2007), where structures grow hierarchically. Small structures form first, then evolve through mergers into large ones. In
the event of a galaxy merger, the less massive galaxy (satellite)
sinks into the gravitational potential of the massive central galaxy
as a result of dynamical friction (Binney \& Tremaine 1987). In our model we
assume that if the mass of the satellite galaxy is comparable to that of the central galaxy, the
merger disrupts the galaxies and resulting in the formation of an elliptical
galaxy (major merger). This event is accompanied by a burst of star
formation as the available cold gas from both progenitors is
transferred to the centre and transformed into stars. Some of this
cold gas reservoir feeds the central SMBH. An additional process of cold gas accretion that contributes to the SMBH mass is the collapse of galaxy discs triggered by dynamical instabilities (Efstathiou, Lake \& Negroponte 1982). When the self-gravity of the galactic disc becomes sufficiently large, the disc equilibrium is 
disrupted resulting in the formation of a bar which enables 
gas to be transferred towards the centre as the disc.
A fraction of that gas is directly fed into the BH through an accretion
disk and powers an AGN (Lynden-Bell 1969), while the rest undergoes star formation. 
For highly efficient accretion
activity, the disc/BH system dominates the energetics of the nucleus,
and the galaxy becomes visible as a quasar. SMBHs at the centres of
galaxies observed in the local Universe are regarded as the remnants
of quasar phases at earlier epochs.
 
The consequence of a minor merger, namely, the accretion of satellites 
of low mass compared to the central galaxy, is usually less dramatic. 
During a minor merger in our model, cold gas and stellar content of the satellites are added to the central galaxy, and a merger-driven starburst may supply fresh gas to the central BH.
 
SMBHs acquire part of their mass merging with other
SMBHs. The formation of a BH-BH binary and the subsequent coalescence
of the BHs is a natural evolutionary stage for a SMBH if the host
galaxy experiences a merger. As the galaxies merge, 
we assume that the SMBHs sink to the centre by dynamical
friction from distant stars or by viscous effects from the surrounding
gas. The transition to a bound binary state after the galaxy merger is
an open issue (Milosavljevi\'{c} \& Merritt 2001). However, it is
believed that once the separation of the two BHs becomes small enough,
gravitational radiation carries away the remaining angular momentum of
the binary. The removal of energy from a SMBH binary due to
gravitational wave emission leads to a gradual shrinkage of the
relative separation of the two members. There is a point where the
eccentricity reaches zero and the orbit circularises. At that time,
the two BHs are very close to each other, and gravitational waves are
emitted copiously. The radiated energy is so large, that two SMBHs,
which are a few AU apart, lose all their potential energy within a
couple of minutes and inevitably coalesce. After the coalescence is
complete the binary enters the
\textit{ringdown} phase, where the merged members settle into a
quiescent remnant hole.  

A third channel for SMBH growth in our model is provided by diffuse gas 
in dark matter haloes undergoing quasi-hydrostatic cooling. 
When a massive halo collapses gas is shock heated out to 
a radius comparable to the
virial radius of the dark matter halo. These haloes have a cooling
time that is longer than the free-fall time of the gas and, thus, the gas settles
into a quasi-static atmosphere surrounding the galaxy rather than
simply falling towards the centre (White \& Frenk 1991). This
atmosphere -- the ``hot halo" regime -- is in pressure supported
hydrostatic equilibrium and extends beyond the virial radius of the
dark matter halo. The galaxy is then supplied with cold gas by 
cooling flows at the disc centre, which also feed the central SMBH.

The formation of massive hot haloes would lead
to the growth of very massive galaxies unless a heating
mechanism regulates the cooling flow. Bower \etal (2006)
and Croton \etal (2006) invoke energy injection
from the central SMBH during the so-called ``radio mode"
feedback (see also de Lucia \etal 2006; Cattaneo \etal 2007; Lagos \etal 2008). 
The radio mode is assumed to occur during the quiescent
accretion of gas from the hydrostatically supported hot halo onto the
SMBH, during which energy from the BH accretion is injected directly
into the hot halo suppressing the cooling flow. In the Bower \etal
(2006) model, the cooling flow stops when the power from the SMBH
is sufficient to offset the rate at which energy is being radiated
away.

Bower \etal (2006) adopt a BH growth model in which during a disc instability 
or galaxy merger, the BH accretes a fixed fraction of the gas that turns into stars
in the burst taking into account processes such as feedback and recycling in 
the galaxy (Malbon \etal 2007). The amount of gas deposited
onto the BH is set by an efficiency factor, which determines the
fraction of the available gas reservoir for star formation that is
accreted by the hole. The value of that parameter is chosen to fit the
normalisation of the local $\MBH-M_{\rm bulge}$ relation.  Hereafter, we refer to the
accretion of cold gas triggered by disc instabilities and galaxy
mergers as ``quasar mode" accretion and to the accretion from
quasi-hydrostatic haloes as ``radio mode" accretion, following the terminology
introduced by Croton \etal (2006).

\subsection{Astrophysical processes affecting BH spin evolution} \label{sec:Spin evolution}

SMBHs are expected to possess angular momentum $ \JBH=|a|G\MBH^2/c$,
where $a$ is the spin parameter, $0\leq |a|\leq1$. The spin has a significant impact in the close
vicinity of the BH. For example, it determines the efficiency for
converting matter into radiation in an accretion disc (Novikov \&
Thorne 1973) and it is believed to influence the formation and
direction of the radio jets in AGN (Blandford \&
Znajek 1977; Macdonald \& Thorne 1982; Begelman, Blandford \& Rees
1984). In addition, it is of special interest in the modelling of gravitational 
waves from BH-BH binaries (Merritt \etal 2005;
Baker \etal 2006a, 2007; Campanelli \etal 2007; Buonanno \etal 2007).

The evolution of spin is closely related to the channels
that contribute to the growth of the BH. Each
mechanism for BH growth is associated with different spin evolution. For
example, accretion of gas that co-rotates with the BH should spin up
the hole (Bardeen 1970), whereas, the merger of two equal-mass 
non-spinning BHs results in a final spin of $0.69$ (Baker \etal 
2006b; Berti \etal 2007, Hinder \etal 2008). We explain below 
how these mechanisms that influence the spin of a BH are 
included in our model.

\subsubsection{Gas accretion} \label{sec:Spin evolution due to accretion}

After the SMBH seed forms at the centre of a galaxy, accretion 
usually initiates the growth era. We assume that an accretion
disc is formed in the equatorial plane of the hole. 
As proposed by Lynden-Bell (1969), gas parcels 
in the disc gradually lose angular momentum due to viscous 
torques exerted by magnetic fields and drift radially inwards 
until they reach the inner edge of the accretion disc. The inner edge 
of the disc is usually taken as the location of the 
last stable orbit (LSO) around the BH. The LSO
is a function of  the hole's angular momentum and can 
be written as (Bardeen \etal 1972):
\begin{equation}
\hrlso\equiv\rlso/\Rg=\{3+Z_2\pm[(3-Z_1)(3+Z_1+2Z_2)]^{1/2}\},
\end{equation}
where the gravitational radius, $\Rg$, is defined as half of the Schwarzschild radius of the BH, $\Rsch=2\Rg=2G\MBH/c^2$, and $Z_1$, $Z_2$ are defined in terms of the spin, $a$, as
\begin{eqnarray}
Z_1&\equiv&1+\left(1-a^2\right)^{1/3}\left[\left(1+a\right)^{1/3}+\left(1- a\right)^{1/3}\right],\nonumber\\
Z_2&\equiv&\left(3 a^2+Z_1^2\right)^{1/2}.
\end{eqnarray}
Note that when the spin parameter has a negative sign the BH is counter-rotating with respect to the orbit of a particle around it. Then, for counter-rotating orbits ($-1 \leqslant a < 0$) $9 \leqslant\hrlso< 6$ and,
for co-rotating orbits ($0 < a \leqslant 1$), $6<\hrlso\leqslant 1$.

An important property is the binding energy of the gas at the LSO, defined as the difference between the rest energy of a gas parcel at infinity and its energy at the LSO as measured by an observer at infinity. If $\tilde{e}$ expresses the energy per unit rest mass, we can define the binding energy as $1-\tilde{e}/c^2$. This provides a simple relation for the accretion efficiency,
\begin{equation}
\epsilon\equiv1-\tilde{e}_{\mathrm{lso}}/c^2=
1-\sqrt{1-\frac{2}{3}\frac{1}{\hrlso}},
\label{efficiency}
\end{equation} 
(Novikov \& Thorne 1973) which corresponds to the fraction of the energy
released by matter spiralling in towards the BH through a succession
of almost circular orbits. Eq. (\ref{efficiency}) shows that accretion
of matter onto slowly rotating BHs has moderate efficiency. For non-rotating BHs ($a=0$,
$\hrlso=6$) this is just $1-\sqrt{8/9}$ or $5.7\%$. However, for
co-rotating matter around rapidly rotating BHs, the efficiency
increases significantly reaching $1-1/\sqrt{3}$ or $42.3\%$ as $
a\rightarrow1$. This sets an upper limit to the efficiency of the
accretion onto BHs, a process obviously much more efficient than
thermonuclear burning.

\begin{figure}
\center
\includegraphics[scale=0.43]{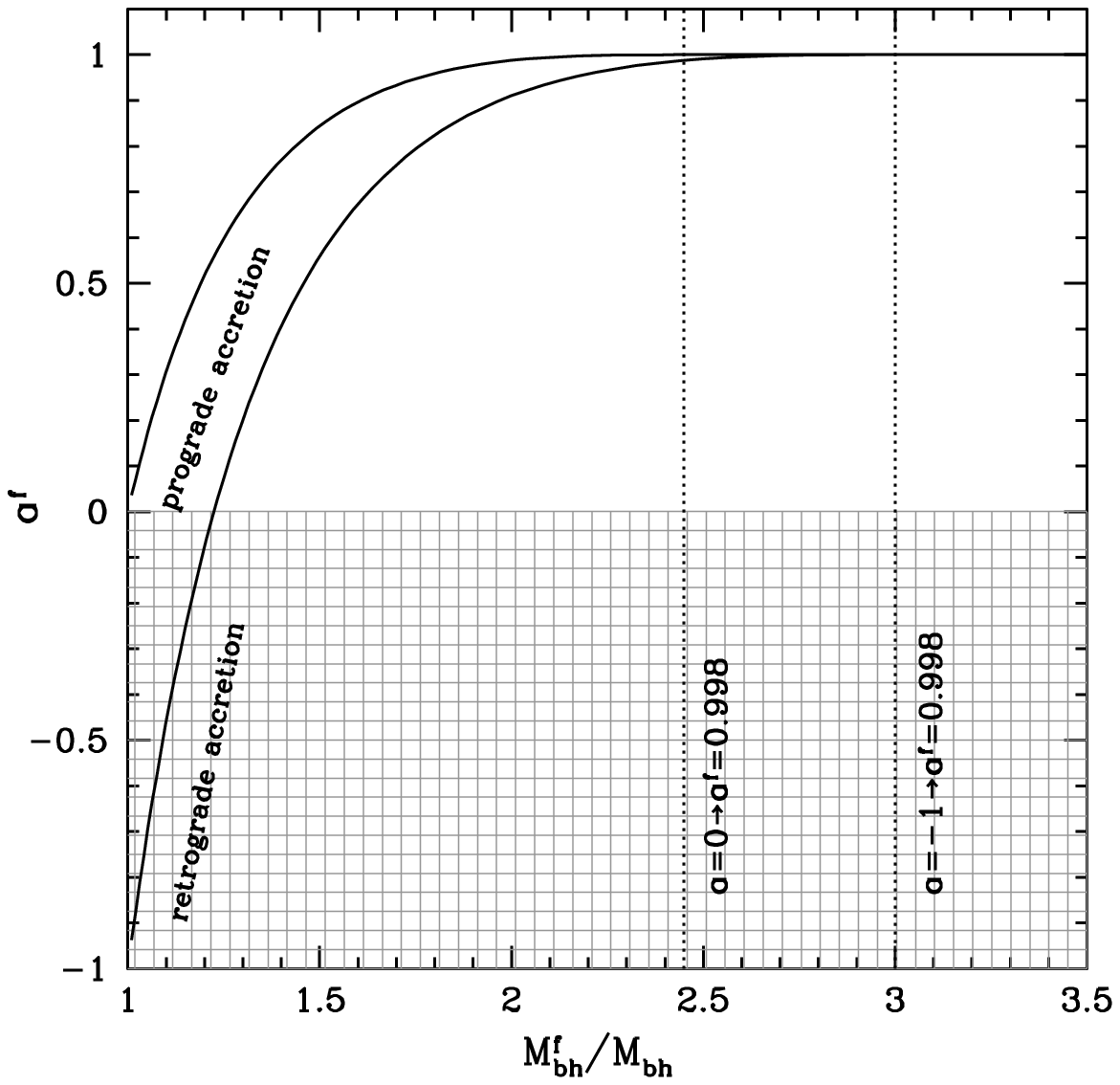}
\caption{The final BH spin after the accretion of gas in a counter- and co-rotating configuration. The dotted lines indicate the final mass needed to spin up a non-rotating and maximally counter-rotating BH to $a^{f}=0.998$. }
\label{bardeen}
\end{figure}

Once the gas reaches the LSO, we assume that it falls directly into the BH. In this way, the accretion carries into the BH the energy per unit mass, $\tilde{e}$, and angular momentum per unit mass, $\tilde{l}$, that the gas has at the LSO. Thus, accretion of a rest mass $\mathrm d M_0$ leads to a change in the hole's total mass, $\MBH$, and angular momentum, $\JBH$, equal to,
\begin{equation}
\mathrm d\MBH=(\tilde{e}_{\mathrm{lso}}/c^2)\mathrm d M_0,\hspace{0.2cm}\mathrm d\JBH=\tilde{l}_{\mathrm{lso}}\mathrm d M_0.
\end{equation}
The change in the hole's spin induced by the accretion of $dM_0$ is governed by the differential equation
\begin{equation} 
\frac{\mathrm d a}{\mathrm d\ln\MBH}=\frac{1}{\MBH}\frac{c^3}{G}\frac{\tl_{\mathrm{lso}}}{\te_{\mathrm{lso}}}-2 a.
\label{spin_change_de}
\end{equation} 
This was integrated by Bardeen (1970) using the explicit expressions for $\te$ and $\tl$, resulting in the following solution,
\begin{equation}
\af =\frac{1}{3}\hrlso^{1/2}\frac{\MBH}{\MBH^{f}}\left[1-\left\{3\hrlso\left(\frac{\MBH}{\MBH^{f}}\right)^2-2\right\}^{1/2}\right]
\label{spin_evolution}
\end{equation}
when $M_{\mathrm{bh}}^{f}/\MBH\leq\hrlso^{1/2}$. Here, $\MBH^f$ and $a^{f}$ are the final mass and spin parameter of the BH. If $M_{\mathrm{bh}}^{f}/\MBH>\hrlso^{1/2}$, then the final spin is always equal to unity. The expression in Eq. (\ref{spin_evolution}) governs the evolution of $ a$ during accretion from an initial state of a co-rotating or counter-rotating disc. According to this, a non-rotating BH will be spun up to a maximum rotation ($ a=1$, $\hrlso=1$) after increasing its mass by $1.44\MBH$ (see Fig.~\ref{bardeen}). For a maximally rotating BH in a counter-rotating accretion disc ($a=-1$, $\hrlso=9$),  an increase in mass of  $2\MBH$ is required in order for the BH to be spun down to $ a=0$ and subsequently spun up to maximum rotation. As implied by Eq. (\ref{spin_evolution}), further accretion onto the BH keeps $a$ equal to unity.

Bardeen's calculations are limited to the process of angular momentum transport between the accreted matter and the BH, without taking into account any other processes. In fact, Thorne (1974) argued that the accretion disc radiates and that some of this radiation will be accreted by the hole. Capture of photons with angular momentum opposite to that of the BH will then produce a counteracting torque that prevents spin up beyond the limiting value of $ a=0.998$ ($\hrlso=1.23$, $\epsilon\simeq0.32$). In addition, axial relativistic outflows of matter in the form of jets may have strong implications for the upper spin limit, since these outflows are accelerated by magnetic fields that are powered by the extraction of the rotational energy of the BH (Hawley \& Krolik 2006; Benson \& Babul 2009).

\subsubsection{Gas accretion through a misaligned
disc}\label{subsec:The case of misaligned accretion discs} 

In the general case where the accretion disc does not lie on the equatorial 
plane of the BH but has a random orientation relative to the angular momentum of 
the BH, the evolution of the spin is a complicated process. 
Following the discussion in King \etal (2005), we assume that a thin disc is 
inclined at some random angle $\theta$ relative to the orientation of the hole's 
angular momentum vector, $\vjh$. We denote the angular momentum of the disc as 
$\vjd$ (see King \etal 2005 and Volonteri \etal 2007 for a discussion of the 
nature of $\vjd$), and define the total angular momentum vector $\vjt$ as 
\begin{equation} \vjt=\vjh+\vjd. \end{equation} 
The angle $\theta$ between 
$\vjh$ and $\vjd$ is defined such that $0\leq\theta\leq\pi$; the values 
$\theta=0$ and $\theta=\pi$ correspond to full alignment and anti-alignment 
respectively. $\vjt$ is a constant vector i.e. has fixed orientation as well
as magnitude for a given accretion event.  Its magnitude is given by 
\begin{equation} 
\Jt^2=\JBH^2+\JD^2+2\JBH\JD\cos\theta. 
\end{equation} 
We further introduce the 
angle $\theta_t$ as the angle between $\vjh$ and $\vjt$.

\begin{figure}
\center
\includegraphics[scale=0.45]{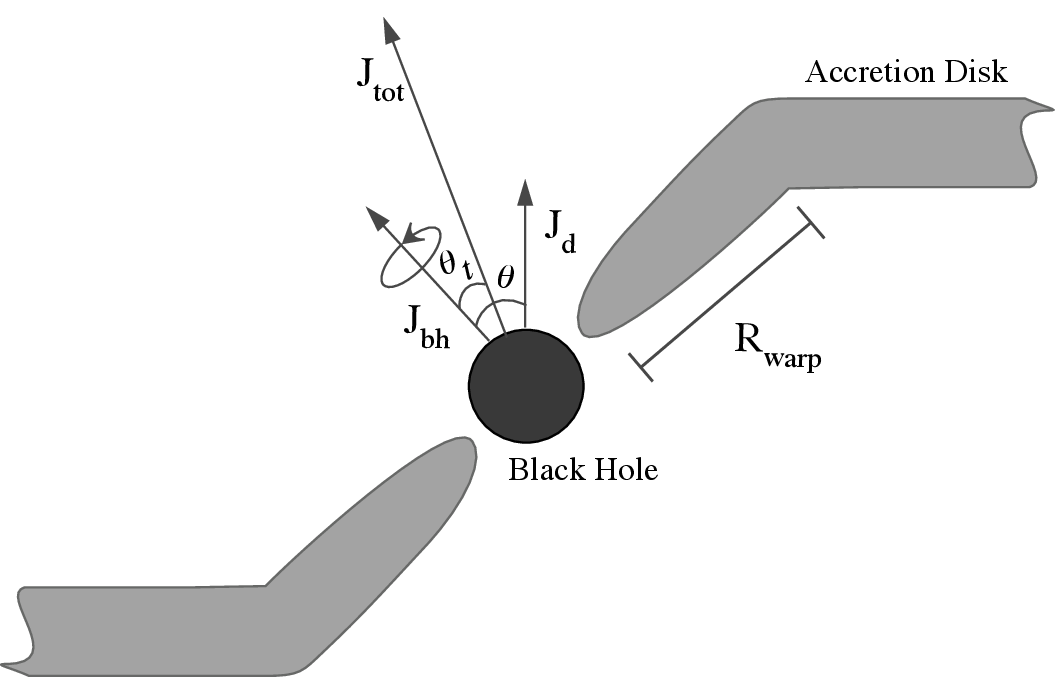}
\caption{Schematic illustration of a warped accretion disc. $\vjh$ is the angular momentum of the BH ($| \vjh|=|a|G\MBH^2/c$), $\vjd$ is the angular momentum of the disc given by Eq.~(\ref{ang_disc}) and $\vjt$ represents the total angular momentum of the system, $\vjh+\vjd$. }
\label{warped_disc}
\end{figure}

When the vectors $\vjh$ and $\vjd$ are misaligned, the tilted orbits of the 
gas parcels in the accretion disc experience a torque due to the Lense--Thirring 
effect, which causes the plane of the accretion disc to precess about the 
rotational axis of the BH (Lense \& Thirring 1918; Wilkins 1972; Bardeen \& 
Petterson, 1975; Scheuer \& Feiler, 1996). If viscosity is strong enough this can 
force the inner  parts of the disc to rotate into the equatorial plane of the hole 
resulting in a ``warped disc" (see Fig.~\ref{warped_disc}).

The Lense--Thirring torque can be expressed as 
\begin{equation}
\frac{\partial \boldsymbol{L}}{\partial t}=\boldsymbol{\Omega_{\mathrm{p}}}\times\boldsymbol{L},
\end{equation}
where $\boldsymbol{L}$ is the angular momentum per unit area of the disc and $\boldsymbol{\Omega_{\mathrm{p}}}$ is the precession rate and is given by
\begin{equation}
\boldsymbol{\Omega_{\mathrm{p}}}=\frac{2G\vjh}{c^2R^3},
\end{equation}
where $R$ is the distance from the BH (Pringle 1992). The precession timescale is therefore defined as,
\begin{equation}
t_{\mathrm{prec}}\equiv\frac{2\pi}{{\Omega}_{\mathrm{p}}(R)},
\end{equation}
which is proportional to $R^{3}$ and thus is much shorter closer to the BH. Other timescales relevant to this problem are the viscous timescales of accretion, $t_{\nu_1}$, and warp propagation, $t_{\nu_2}$,
\begin{equation}
t_{\nu_{1,2}}\equiv\frac{R^2}{\nu_{1,2}(R)}. 
\end{equation}
 Here, $\nu_1$ and $\nu_2$ are the kinematic viscosities acting on velocity gradients parallel and normal to the plane of the disc respectively. The balance between the timescales $t_{\mathrm{prec}}$ and $t_{\nu_2}$ determines whether the Lense -- Thirring torque is able to align the inner disc with the spin axis. The condition for alignment is $t_{\mathrm{prec}}\lesssim t_{\nu_2}$ (Natarajan \& Armitage 1999). In other words, the disc will be aligned with the spin at radii where the precession timescale is much shorter than the timescale of radial diffusion of the warp. The characteristic radius, $\Rw$, of the aligned part of the disc follows from the condition $t_{\mathrm{prec}}=t_{\nu_2}$.
  
The evolution of $\vjh$ and $\vjd$ during the precession is determined by the conditions (King \etal 2005):
\begin{equation}
\frac{\mathrm d}{\mathrm d t}\JBH^2=0,\;\;\;\frac{\mathrm d}{\mathrm d t}\JD^2\leq0\;\;\;\mathrm{and}\;\;\;\frac{\mathrm d}{\mathrm d t}\cos\theta_t\geq0.
\end{equation}
As can be inferred by the first two conditions, the magnitude of $\vjh$ remains constant while the angle $\theta_t$ decreases with time, which implies that $\vjh$ always aligns with $\vjt$. In contrast, $\vjd$ decreases in magnitude as $\vjh$ aligns, which is to be expected, since the total angular momentum has to remain constant. 

The end result of the Lense -- Thirring precession is a BH which is aligned or anti-aligned with the surrounding accretion disc. The expression for the magnitude of $\vjt$ allows us to examine the final configuration of the system. Obviously, if $\JBH^2>\Jt^2$ anti-alignment occurs, which requires
\begin{equation}
\cos\theta<-\frac{\JD}{2\JBH}.
\label{alignment_criterion}
\end{equation}
Hence, a BH with $\theta>\pi/2$ {\em and} $2\JBH>\JD$, 
eventually anti-aligns with the accretion disc. This means that if $\JD>2\JBH$ then 
alignment always happens as the disc angular momentum completely overwhelms 
that of the BH.

Further investigation of the warped discs requires knowledge of the accretion flow
properties. We use the thin disc solution of Shakura \& Sunyaev (1973) for our analysis. In the standard Shakura-Sunyaev disc model, the analytic expression for the warp radius depends on the values of $\JBH$, $\MBH$, and the accretion rate $\dot{M}$ of the BH. In terms of the Schwarzschild radius, $\Rw$ is written as (Volonteri \etal 2007)
\begin{eqnarray}
\frac{\Rw}{\Rsch}=3.6\times10^{3} a^{5/8}\left(\frac{\MBH}{10^8~\Msun}\right)^{1/8}\lambda^{-1/4}\hspace{1.5cm}\nonumber\\
\times\left(\frac{\nu_2}{\nu_1}\right)^{-5/8}\alpha^{-1/2}.
\end{eqnarray}
Here $\alpha$ is the Shakura-Sunyaev viscosity parameter and  $\lambda=L/{L_{\mathrm{Edd}}}$ is the Eddington ratio. The accretion luminosity, $L$, and Eddington luminosity, $L_{\mathrm{Edd}}$, are defined as
\begin{equation}
L=\epsilon\dot{M}c^2,
\end{equation}
with $\epsilon$ denoting the accretion efficiency, and
\begin{equation}
L_{\mathrm{Edd}}=\frac{4\pi G\MBH c}{\kappa}=1.4\times10^{46}\left(\frac{\MBH}{10^8~\Msun}\right)~\mathrm{erg~s}^{-1},
\end{equation}
where $\kappa\sim0.3$~cm$^2$g$^{-1}$ is the electron scattering opacity. Associated to the Eddington luminosity is an accretion rate expressed as $\dot{M}_{\mathrm{Edd}}=L_{\mathrm{Edd}}/\epsilon c^2$. This is the accretion rate for which the black hole radiates at the Eddington luminosity. It is then convenient to express the physical accretion rate in units of the Eddington accretion rate, $\dot{m}\equiv\dot{M}/\dot{M}_{\mathrm{Edd}}$. We note that the accretion efficiency in $\dot{M}_{\mathrm{Edd}}$ is spin dependent; however, we chose to keep $\epsilon=0.1$ when calculating $\dot{M}_{\mathrm{Edd}}$ for simplicity.

Given that the accretion is characterised by a timescale $t_{\nu_1}$, the mass of the disc inside the radius $\Rw$ is
\begin{equation}
\Md(\Rw)=\dot{M}t_{\nu_1}(\Rw),
\end{equation}
where the accretion timescale is given by
\begin{eqnarray}
t_{\nu_1}=\frac{\Rw^2}{\nu_1}=3\times10^6~a^{7/8}\left(\frac{\MBH}{10^8~\Msun}\right)^{11/8}\hspace{1.5cm}\nonumber\\
\times\lambda^{-3/4}\left(\frac{\nu_2}{\nu_1}\right)^{-7/8}\alpha^{-3/2}~\mathrm{yr}.
\end{eqnarray}
We can write the total angular momentum, $\JD$, passing through $\Rw$ as, 
\begin{equation}
\JD(\Rw)\lesssim M_d(\Rw)(G\MBH\Rw)^{1/2}.
\label{ang_disc}
\end{equation}
In terms of the anti-alignment criterion this gives,  
\begin{eqnarray}
\frac{\JD}{2\JBH}=\frac{\Md}{a\MBH}\left(\frac{\Rw}{\Rsch}\right)^{1/2}\hspace{3.5cm}\nonumber\\
=10^{-9}\lambda\left(\frac{t_{\nu_1}}{1~\mathrm{yr}}\right)\left(\frac{\Rw}{\Rsch}\right)^{1/2}a^{-1}.
\end{eqnarray}
Evaluation of this quantity determines whether or not the anti-alignment condition in a misaligned disc is satisfied. 

\subsubsection{The case of self-gravity limited discs}\label{subsec:Self-gravity limited accretion discs}

The BH growth process in AGN environments involves vast amounts of accreted gas, 
often comparable to the initial mass of the accreting hole. This amount of mass 
settles onto the accretion disc around the hole, which often extends to several 
thousands of gravitational radii. It is usually assumed that the available mass 
fuel, $\Macc$, is consumed in a single accretion episode, thus providing a 
supply of constant angular momentum (Volonteri \etal 2007). In this case, the 
amount of mass consumed is enough to spin up the hole up to $a=0.998$, even if 
the BH initially was maximally spinning in a counter-rotating direction. 
Inevitably, repeated accretion episodes during major galaxy mergers act to spin 
up the BH to maximum rotation.

Recently, however, King \etal (2008) argued that the end result of the accretion 
growth channel might be completely different if we take into account the fact 
that an accretion disc becomes self-gravitating at some radius, $\Rsg$, where 
its mass exceeds $\Msg\sim(H/R)\MBH$. As a result, the mass of the disc is 
limited by its self-gravity to $\Delta M_{\mathrm{episode}}\ll\Macc$, which 
gives rise to a series of $N\simeq\Macc/\Delta M_{\mathrm{episode}}$ well 
separated accretion episodes. King \etal suggest that these accretion episodes 
are randomly oriented around the BH, an assumption supported by observations 
indicating that there is no apparent relation between the accretion disc (or radio 
jet) orientation and the host galaxy disc (Nagar \& Wilson 1999; Kinney \etal 
2000). They further argue that this could be the result of intense star formation 
outside $\Rsg$ randomising the input gas direction, which could provide a 
qualitative explanation for the ring of stars seen in the near vicinity of the central 
BH in the Milky Way (Genzel \etal 2003).

The much smaller angular momentum associated with each accretion episode 
means that, in general, $\JD<2\JBH$, so in the chaotic accretion model
counteralignment occurs in a fraction (King \etal 
2005) 
\begin{equation} f=\frac{1}{2}\left(1-\frac{\JD}{2\JBH}\right) \simeq 1/2
\label{counter_align_fraction} 
\end{equation} 
of the accretion episodes.  Therefore, counter- and co-alignment are
equally likely outcomes of the Lense -- Thirring effect.  However,
accretion of gas in a counter-rotating disc is more efficient in
spinning up the BH since the gas is being dumped onto the BH from a
larger distance and, thus, carries more angular momentum into the
BH. Hence, a succession of counter-aligned and co-aligned accretion
episodes with equal frequency should systematically spin down the BH,
resulting in a global spin distribution oscillating around zero.

The physics of self-gravitating accretion discs is described by
Pringle (1981). Briefly, the criterion that the self-gravity of a
disc be negligible is the requirement that the gravitational force
along the $\hat{z}$ direction be dominated by the central BH. This
can be expressed as the surface density of the disc, $\Sigma$, being
negligible compared to the quantity $\MBH H/R^3$, or in terms of the
disc mass, \begin{equation} \Md(<R)\ll\frac{H}{R}\MBH.
\end{equation} The self-gravity becomes marginally important when $\Md\simeq 
(H/R)\MBH$, which defines the self-gravity radius, $\Rsg$, given by \begin{eqnarray} 
\frac{\Rsg}{\Rsch}=1.5\times10^{3}~\epsilon^{8/27}\negthinspace\left(\frac{\MBH}
{10^{8}~\Msun}\right)^{-26/27}\hspace{-0.3cm}\lambda^{-8/27} \alpha^{14/27}. 
\end{eqnarray}

The semi-thickness of the disc in the Shakura-Sunyaev model is given by,
\begin{eqnarray}
\frac{H}{R}=1.36\times10^{-3}\epsilon^{-1/5}\left(\frac{\MBH}{10^{8}~\Msun}\right)^{-1/10}\lambda^{1/5}\hspace{1.5cm}\nonumber\\
\times\left(\frac{R}{\Rsch}\right)^{1/20}\alpha^{-1/10}.
\label{semi_thickness}
\end{eqnarray}
Hence, by replacing $R$ with $\Rsg$ in expression (\ref{semi_thickness}), we obtain an analytic expression for the disc mass within $\Rsg$,
\begin{eqnarray}
M_{\mathrm{sg}}=\frac{H}{R}\MBH\Big |_{R=R_{\mathrm{sg}}}=2.13\times10^{5}~\epsilon^{-5/27}\hspace{1.2cm}\nonumber\\
\times\left(\frac{\MBH}{10^{8}~\Msun}\right)^{23/27}
\lambda^{5/27}~
\alpha^{-2/17}\; \Msun.
\label{msg}
\end{eqnarray}
As we described in Section \ref{subsec:The case of misaligned accretion discs}, once the disc forms with a non-zero misalignment angle about the spin axis of the hole, it will be subject to the Lense -- Thirring precession. If the radius, $\Rsg$, is greater than $\Rw$, then
\begin{equation}
\JD=\JD(<\Rw).
\end{equation}
If, however, $\Rsg<\Rw$, then the entire disc will be subject to Lense -- Thirring precession, thus aligning itself in the equatorial plane of the BH.

\subsubsection{BH-BH binary coalescence} \label{sec:Spin evolution due to mergers}

We now examine the evolution of spin during BH binary
coalescence. During the merger of the two binary members, the smaller
BH adds its spin and orbital angular momentum at the LSO to the spin
of the larger one. Even if the progenitor holes do not possess any
orbital angular momentum, the final remnant will preserve the residual orbital
momentum of the binary (i.e. the angular momentum that has not been
radiated away). Thus, the remnant will always be a Kerr BH.
 
Recent breakthroughs in numerical relativity have provided robust
simulations of BH mergers by solving directly the Einstein equations
in a fully relativistic frame (Campanelli \etal 2007; Herrmann \etal
2007; Marronetti \etal 2007). These simulations have explored the
parameter space for different configurations of initial masses and
spins, allowing accurate measurements of the final spin. For example,
a merger of two equal mass BHs both with $a=0$ in a circular orbit
results in a Kerr BH with $a^{f}\simeq0.69$ (Baker \etal 2006b). 
This a robust prediction and is also valid for eccentric orbits with
eccentricities smaller than 0.4 (Hinder \etal 2008).

Analytic fits extend the predictions for $a^{f}$ to the entire 
space of parameters and reproduce closely all
the available numerical data. In the case where the masses are
unequal but the spins are zero, the final spin can be estimated by the analytic expression
\begin{equation} 
a^{f}\simeq2\sqrt{3}\frac{q}{(1+q)^2}-2.029\frac{q^2}{(1+q)^4},
\end{equation}
where $q=M_2/M_1\leqslant1$ is the mass ratio of the progenitor BHs (Berti \etal 2007). 

A more general analytic fitting formula for predicting the final spin of any given binary configuration has been provided by Rezzolla \etal (2008a,b). In their analysis, they assume that the final spin parameter vector can be expressed as 	
\begin{equation} 
\boldsymbol{a}^{f}=\frac{1}{(1+q)^2}(\boldsymbol{a_1}+ \boldsymbol{a_2}q^2+\boldsymbol{\ell}q),
\end{equation}
where $\boldsymbol{a_{1,2}}=c\boldsymbol{\mathrm{J}}^{1,2}_{\mathrm{bh}}/(GM^2_{1,2})$ and $\boldsymbol{\ell}= \boldsymbol{\ell'}/(M_1 M_2)$, with $\boldsymbol{\ell}'$ defining the difference between the orbital angular momentum, $\boldsymbol{l}$, when the binary is widely separated, and the angular momentum radiated away in gravitational waves before the merger, $\boldsymbol{j}_{\mathrm{rad}}$, namely,
\begin{equation}
\boldsymbol{\ell'}=\boldsymbol{l}-\boldsymbol{j}_{\mathrm{rad}}.
\end{equation}
The vector $\boldsymbol{\ell'}$ is taken to be parallel to the orbital momentum vector throughout the evolution of the binary, an assumption which is not strictly valid since the system could radiate away angular momentum in a non-symmetric way. However, the error introduced by this assumption is relatively small for the binary configurations studied in the simulations of Rezzolla \etal A further assumption is that the mass radiated away in gravitational waves is negligible, as it accounts for only a small fraction ($5-7\%$) of the total mass-energy of the binary configurations analysed.
Under these assumptions, Rezzolla \etal (2008a,b) proposed an analytic expression for the magnitude of the final spin, given by
\begin{eqnarray}
\vert\boldsymbol{a}^f\vert=\frac{1}{(1+q)^2}\Big[ \vta{1}^2 + \vta{2}^2 q^4+
 2 \vert{\boldsymbol{a}_2}\vert\vert{\boldsymbol{a}_1}\vert q^2 \cos \varphi \hspace{.6cm}\nonumber\\  
+2\left(\vert{\boldsymbol{a}_1}\vert\cos \vartheta + \vert{\boldsymbol{a}_2}\vert q^2  \cos \xi
\right) \boldsymbol{\ell}{q}+ \boldsymbol{\ell}^2 q^2
\Big]^{1/2},
\label{rezzolla:final_spin}
\end{eqnarray}
with the cosine angles $\varphi,\vartheta$ and $\xi$ defined as
\begin{equation}
\label{cosines}
\cos \varphi \equiv
{\boldsymbol{\hat{a}}_1\cdot\boldsymbol{\hat{a}}_2},
\hspace{0.3cm}
\cos \vartheta \equiv
 \boldsymbol{\hat a}_1\cdot\boldsymbol{\hat{{\ell}}},
\hspace{0.3cm}
\cos \xi \equiv
\boldsymbol{\hat{a}}_2\cdot\boldsymbol{\hat{{\ell}}}.
\end{equation}
The norm of $\boldsymbol{\ell}$ is given by,
\begin{eqnarray}
&&
\vtl= \frac{s_4}{(1+q^2)^2} \left(\vta{1}^2 + \vta{2}^2 q^4 + 2 \vta{1} \vta{2} q^2 \cos\varphi\right)\nonumber \\
&& \hspace{0.8cm} +\left(\frac{s_5 \mu + t_0 + 2}{1+q^2}\right)\left(\vta{1}\cos\vartheta + \vta{2} q^2 \cos\xi\right)\nonumber \\
&& \hspace{0.8cm} +2\sqrt{3}+ t_2 \mu + t_3 \mu^2.
\label{ang_momentum}
\end{eqnarray}
Here $\mu$ expresses the symmetric mass ratio, $\mu=q/(1+q)^2=M_1M_2/(M_1+M_2)^2$, and the coefficients take the values $s_4 =-0.129$, $s_5 = -0.384$, $t_0 = -2.686$,
$t_2 = -3.454$, $t_3 = 2.353$. The final spin as given by Eq. (\ref{rezzolla:final_spin}) is in good agreement with numerical data, with residuals of less than $3\%$.

\subsection{Simulating the spin evolution}\label{sec:Simulating the spin evolution}

We finally describe our method for calculating the evolution of BH
spin. During the formation history of the BHs we take into account
spin changes due to the accretion of gas and mergers, as described in
Section \ref{sec:Spin evolution}. The initial population of seeds is
assumed to be non-rotating. These seeds grow through accretion of gas
during disc instabilities, galaxy mergers, and radio mode accretion and
through BH-BH mergers.

\subsubsection{Thin discs}
To model the physics of the accreted gas, we assume that when  $\dot{m}\geqslant0.01$ the gas forms an accretion disc around the BH whose physics is described by the standard Shakura-Sunyaev disc model. We assume that the gas available to be fed into the BH after a galaxy merger or the collapse of a dynamically unstable disc is accreted over a timescale proportional to the dynamical timescale of the galactic bulge that hosts the BH. A fixed proportionality factor of 4 is used in this analysis resulting in typical accretion timescales of the order of $10^{7}-10^{8}$~yr. Note that, this factor is $\sim10$ longer than the one chosen by Malbon \etal (2007). The new value is introduced in order to match the local quasar luminosity function and does not affect the galaxy-formation model. The timescale for accretion during the radio mode is computed directly by the galaxy formation model and depends on the cooling timescale of the gas in the hot halo.

The accretion disc is assumed to be randomly oriented relative to the
spin axis of the hole. If $\theta\neq0^{\circ},180^{\circ}$, we assume
that the disc precesses around the BH as described in Section
\ref{subsec:The case of misaligned accretion discs}. In brief, for a
given BH mass, $\MBH$, we determine the numerical values of the warp
parameters, $\Rw$, $\Md(\Rw)$, and the angular momenta $\JD(\Rw)$ and
$\JBH$.  The ratio $\JD/2\JBH$ allows us, through the criterion in
Eq. (\ref{alignment_criterion}), to check if anti-alignment or
alignment occurs. We then allow the BH to accrete an amount of gas
equal to $\Md(\Rw)$ and evolve the spin using
Eq. (\ref{spin_evolution}). Finally, we update the mass of the BH and
estimate the hole's new spin and the disc precession by calculating
the new angles $\theta$ and $\theta_t$. This process is repeated
until: (a) the disc aligns or counteraligns ($\theta=0^{\circ}$ or
$180^{\circ}$) with the hole's spin, in which case we just consume the
rest of the available gas and evolve the spin according to
Eq. (\ref{spin_evolution}), or (b) the accretion disc is entirely
consumed, without being able to align or counteralign itself with the
spin. In almost all cases we find that the accretion disc ultimately
aligns (or anti-aligns) itself on the equatorial plane of the BH.

In addition to the case where the entire gas reservoir is consumed 
within one accretion episode with constant angular momentum orientation
 (prolonged accretion), we test the model 
proposed by King \etal (2008) in which the mass of the accretion disc is 
limited by its self-gravity (chaotic accretion). We repeat the same steps as 
before. However, in this case $\Md=\Msg$. Once $\Msg$ is consumed, we 
update the mass of the hole and determine the mass of the new disc, 
$\Msg'$, computed from the updated BH mass. At this point we make the 
assumption that the disc retains no memory of the angular momentum 
of the initial flow and that it settles into a random orientation around the BH with 
$\Md=\Msg'$. We then add the next $\Msg$ until the disc is consumed. 

It is important to clarify at this point that the two accretion models considered 
here serve as a means for studying the effect of accretion on BH spin. The 
forthcoming analysis that follows and results do not rely on any specific
model for the physical processes that drive the gas flows towards the 
vicinity of the BH. Several mechanisms for feeding the SMBH with 
gas could result to similar spin distributions to the chaotic accretion case 
considered here (see for example, Heller \etal 2001 for the 
properties of non self-gravitating gaseous bars in barred disc galaxies and Hobbs 
\etal 2010 for an investigation of the effect of supernova-driven turbulence on 
the fuelling of SMBHs in galactic bulges). Here, our main goal is to 
explore the effect of different SMBH spin distributions on properties of the host galaxy.
 
\subsubsection{ADAFs}

Many of the accretion events in our simulations are characterised by low, 
sub-Eddington, accretion rates, with $\dot{m}\leqslant0.01$. For such low accretion
rates, the gas flow has low density and is unable to cool efficiently
since radiative cooling does not balance the energy generated by
viscosity. Thus, the viscous energy is trapped in the gas as entropy
and ultimately advected into the hole.  This type of accretion is
known as an ADAF (Rees \etal 1982; Narayan \& Yi 1994; Abramowicz
\etal 1995).

ADAFs have a number of distinct properties that some of them will be
essential for the analysis in later sections (see Sections 5 and 6). 
For example, for an ADAF around a BH, only a fraction of the standard accretion
luminosity, $L=\epsilon\dot{M}c^2$, is emitted as radiation. The
remainder of the viscously dissipated energy is stored in the gas as
entropy, resulting in hot flows with almost virial temperatures. We
note that, as shown by Ichimaru (1977), the ions and the electrons in
an ADAF are not thermally coupled and, thus, reach different
temperatures. This two-temperature virialised plasma flow is optically
thin and, for high viscosity parameters ($\alpha\sim0.2-0.3$), it
acquires a quasi-spherical geometry around the BH ($H\sim R$), which
resembles spherical Bondi accretion. However, despite the geometrical
similarity, the dynamics of the ADAF are fundamentally different,
since the accretion is entirely due to dissipation via viscous forces
rather than gravity.

In general, the accretion flow can have a rather complicated
structure. For example, it is possible that a thin disc dominates the
geometry of the outer parts, then switches to an ADAF as the flow
approaches the BH (Esin \etal 1997). The transition to the thick flow
depends on the accretion rate and usually occurs closer to the BH for
higher accretion rates. Such a complex disc model for describing the
accretion flows during the radio mode is, however, beyond the scope of
this paper. Here, we assume a simple configuration of a
quasi-spherical corona around the hole with any transition to a thin
disc occurring far from the BH. Precession effects due to misalignment
between the disc and the BH are modelled as in the thin disc case.

\subsubsection{Binary mergers}

Finally, we consider the spin changes due to binary coalescence. 
Following the merger of two galaxies in \texttt{GALFORM}\normalsize, 
each harbouring a SMBH, the one hosted by the satellite 
sinks towards the SMBH of the central galaxy and eventually form 
a binary. In our model, BH mergers tend to occur in gas-poor 
environments, and thus, we assume that torques from accreting flows 
that might be present during the formation of the binary do not influence 
the orientation of the BH spins prior the merger. We therefore take the 
BH spin vectors in Eq. (\ref{rezzolla:final_spin}) to be randomly oriented 
relative to each other and the orbital angular momentum. In addition, 
we assume that all binaries will ultimately merge on a very short timescale, 
effectively right after the host galaxies merger and induced accretion have
been completed.

During the merger, the smaller BH plunges into the larger 
one carrying along its angular momentum at the LSO. At this point 
we neglect any mass loss due to gravitational wave emission, as it 
corresponds to a small fraction of the total mass-energy of the system. 
In addition, we do not take into account the effects of gravitational 
recoil  due to asymmetric emission of gravitational waves in binary 
mergers. In our model, mergers with recoil velocities as high
as  $\upsilon_{\mathrm{recoil}}\sim500$~km~s$^{-1}$ (Libeskind \etal 
2006) could displace BH remnants. However, we assume that dynamical friction
relocates them in the galactic centre. Therefore, we do not 
expect gravitational recoil to have a significant impact 
on the spin distribution of the remnant BHs.
The spin of the final remnant is finally estimated using the fitting formulae in 
Eqns. (\ref{rezzolla:final_spin}) \& (\ref{ang_momentum}), 
as described in Section \ref{sec:Spin evolution due to mergers}. 

\section{Evolution of BH mass and spin} \label{sec:evolution of mass and spin} 

\subsection{Evolution of BH mass}\label{sec:mass evolution}

\begin{figure}
\center \includegraphics[scale=0.43]{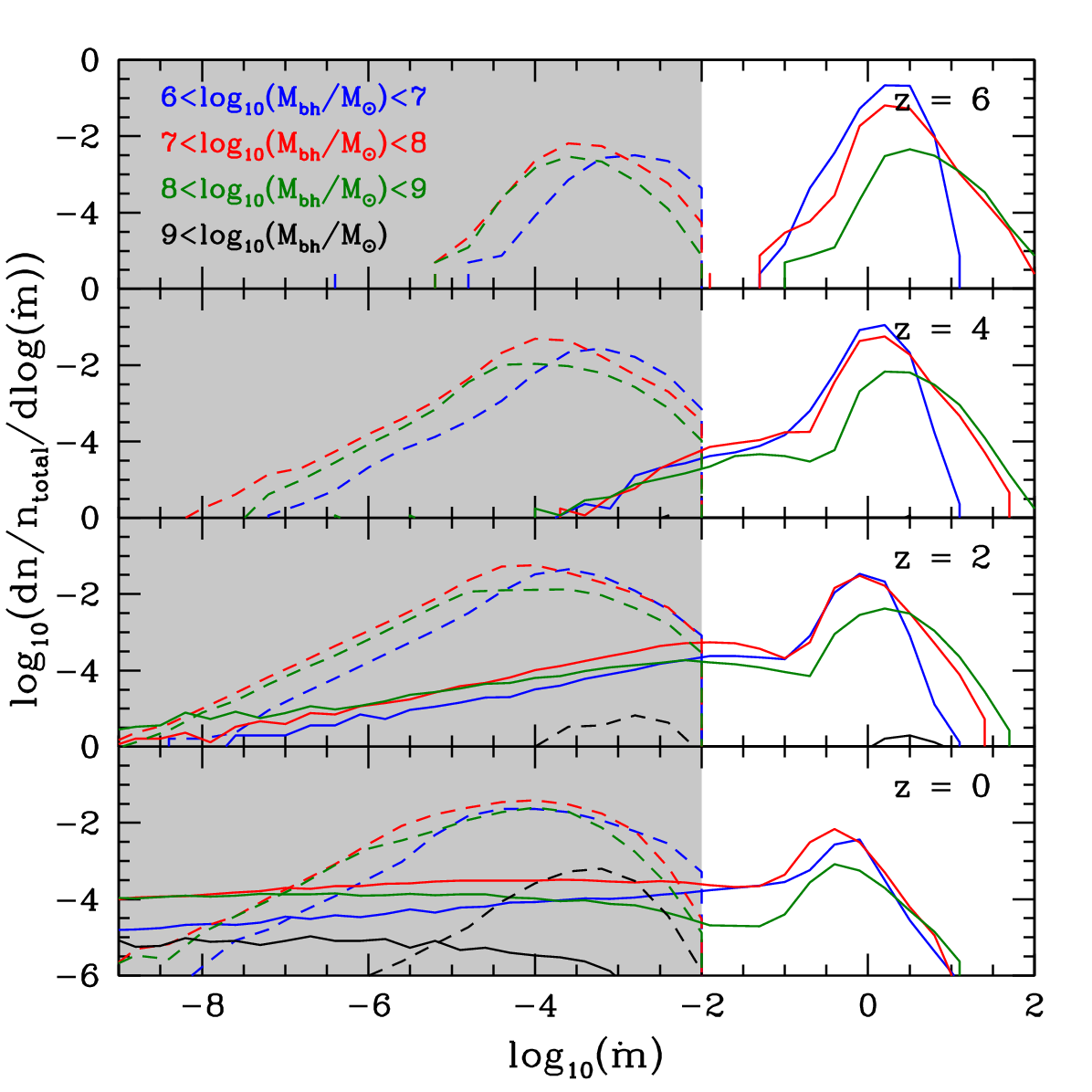}
\caption{The distribution of accretion rates for selected BH mass
ranges at various redshifts. Line styles represent two different
accretion modes: solid lines for the quasar mode and dashed lines
for the radio mode. The shaded area represents the regime where the
accretion flow is described by an ADAF. The accretion rates in the
radio mode are truncated at $\dot{m}=0.01$, in
accordance with our model of AGN feedback. In the present Universe,
the quasar mode accretion peaks at $\dot{m}\simeq0.5$. However, 
the peak shifts to higher values at higher redshifts, indicating that 
accretion was more efficient in the past.}
\label{mdot}
\end{figure}
\begin{figure}
\center \includegraphics[scale=0.43]{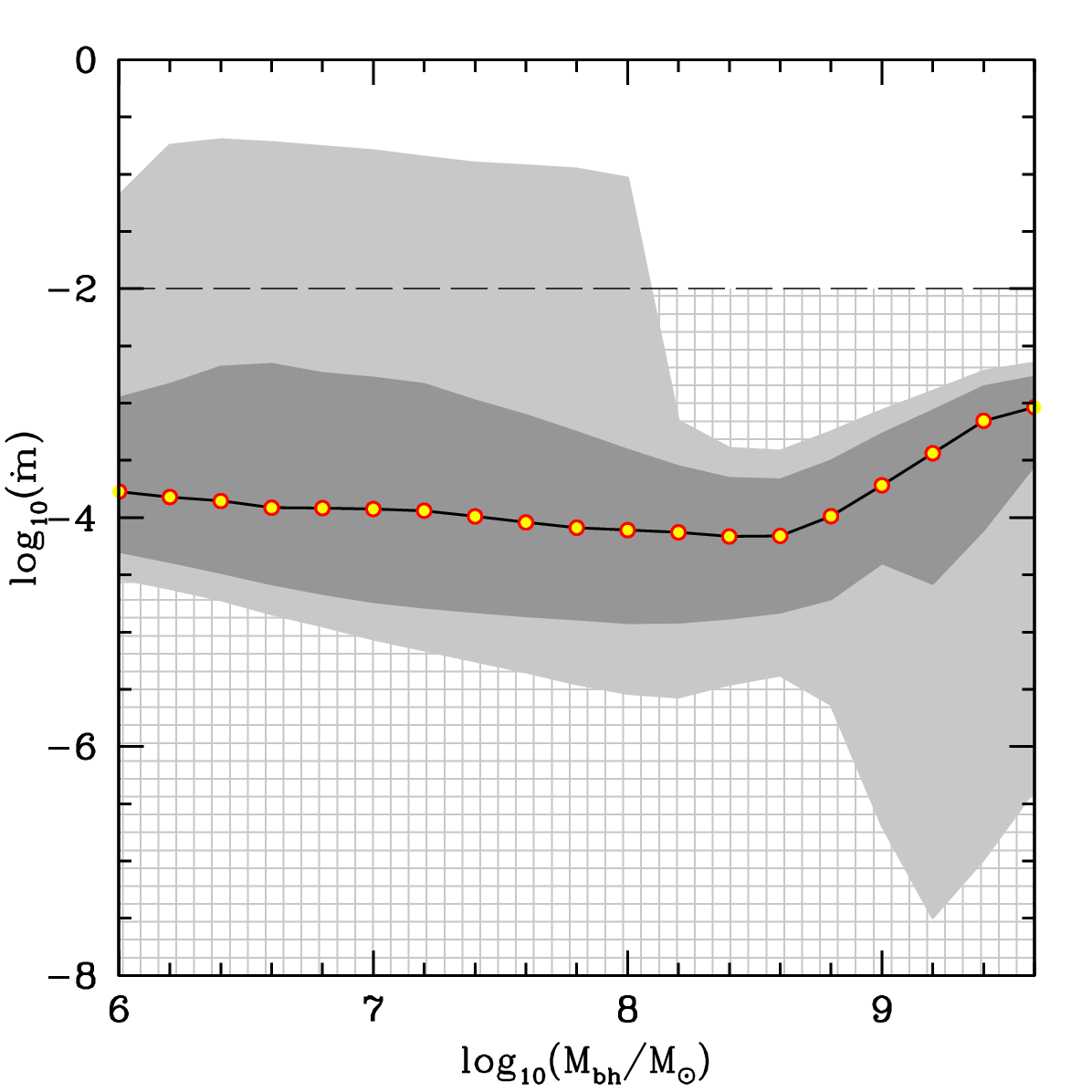}
\caption{The correlation of accretion rate with BH 
mass at redshift zero. The median (red-yellow points), $20-80$
percentiles (dark grey) and $5-95$ percentiles (light grey) are
shown. The shaded background indicates the region were the accretion
disc is modelled as an ADAF. Only lower mass BHs ($<10^8~\Msun$)
i.e. those hosted by lower mass galaxies (typically spirals) have
enough gas left to trigger accretion via a thin disc
($\dot{m}>0.01$). More massive BHs have little cold gas and
undergo radio mode accretion which (by construction) has $\dot{m}<0.01$.}
\label{mbh_mdot}
\end{figure}
\begin{figure}
\center
\includegraphics[scale=0.43]{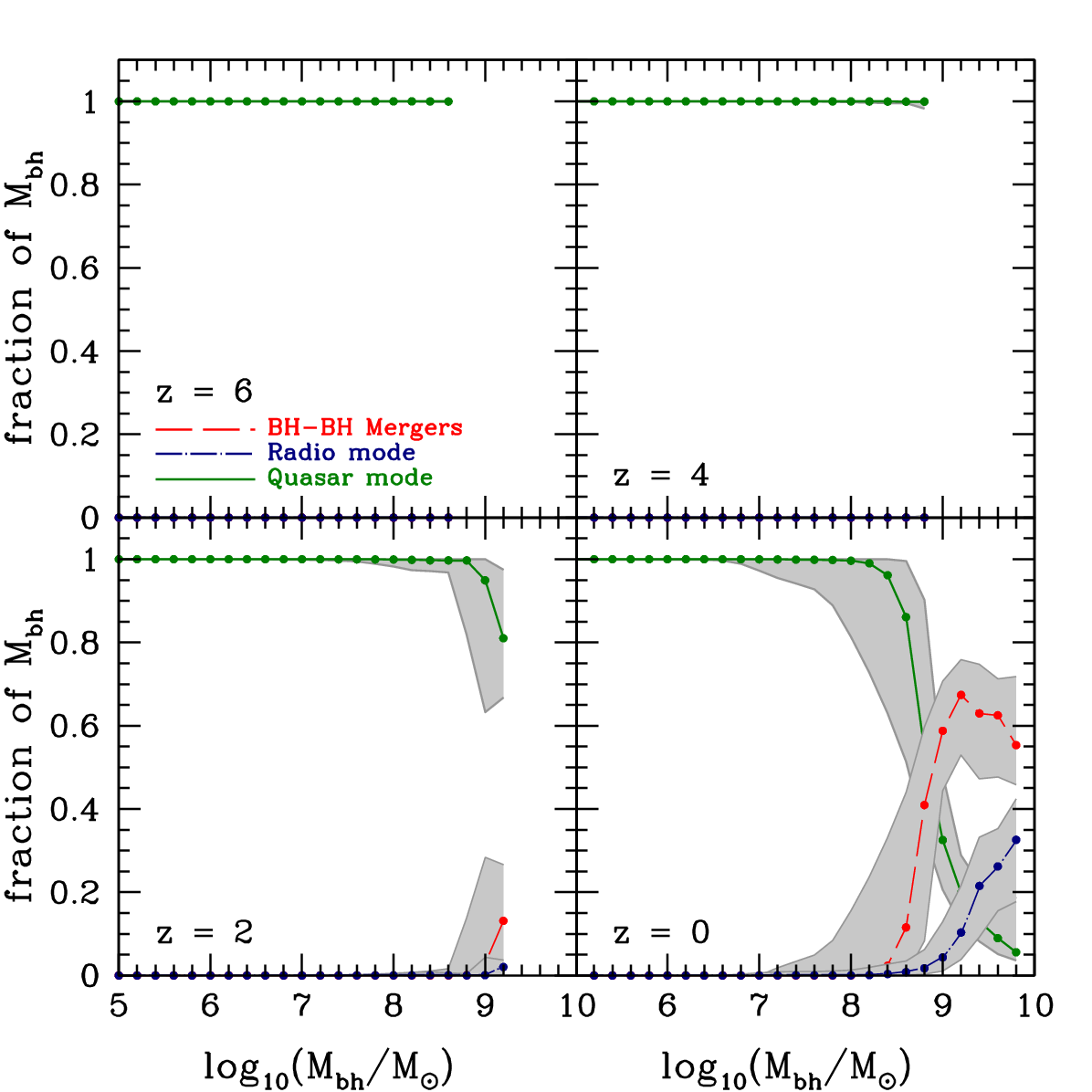}
\caption{Contribution to the final BH mass at different redshifts from
gas accretion in quasar mode (solid green line), radio mode (dashed
blue line) and BH-BH mergers (dashed-dotted red line) in the updated Bower \etal (2006) model. The lines represent the medians and the shaded region the 20-80
percentiles of the distributions. The dominant channel of BH growth for BHs with
$\MBH\lesssim10^{8}~\Msun$ is the quasar mode. Above that mass,
mergers and radio mode become the dominant growth channels.}
\label{growth_channels}
\end{figure}

We now briefly present our main predictions for BH growth, using our
updates (see Section 2) to the model developed by Bower \etal (2006)
and Malbon \etal (2007). In Fig.~\ref{mdot} we show the fraction of
all BHs that accrete at a 
given rate in the quasar and radio modes, at different redshifts.  All
BHs that accrete in the radio mode during the redshift bin are
included, but the quasar mode is episodic so we only
include objects which experience a starburst within a given redshift
bin. Our updated parameters give a longer duration for the accretion
episode in the quasar mode ($4$ times rather than $0.5$ times the
dynamical timescale of the bulge) than assumed in Malbon \etal 2007. This results
in the majority of BHs with mass $10^6-10^8~\Msun$ which accrete in
the quasar mode at $z=0$ having $\dot{m}\simeq 0.5$, as observed (Heckman
\etal 2004). At higher redshift the distributions shift to higher
mass accretion rates, with a mean $\dot{m}\simeq3$ at $z=6$. This
implies that in a hierarchical universe, accretion of gas was more
efficient at early epochs, so the rate of growth of BHs is faster in
the past. 
\begin{figure}
\center
\includegraphics[scale=0.49]{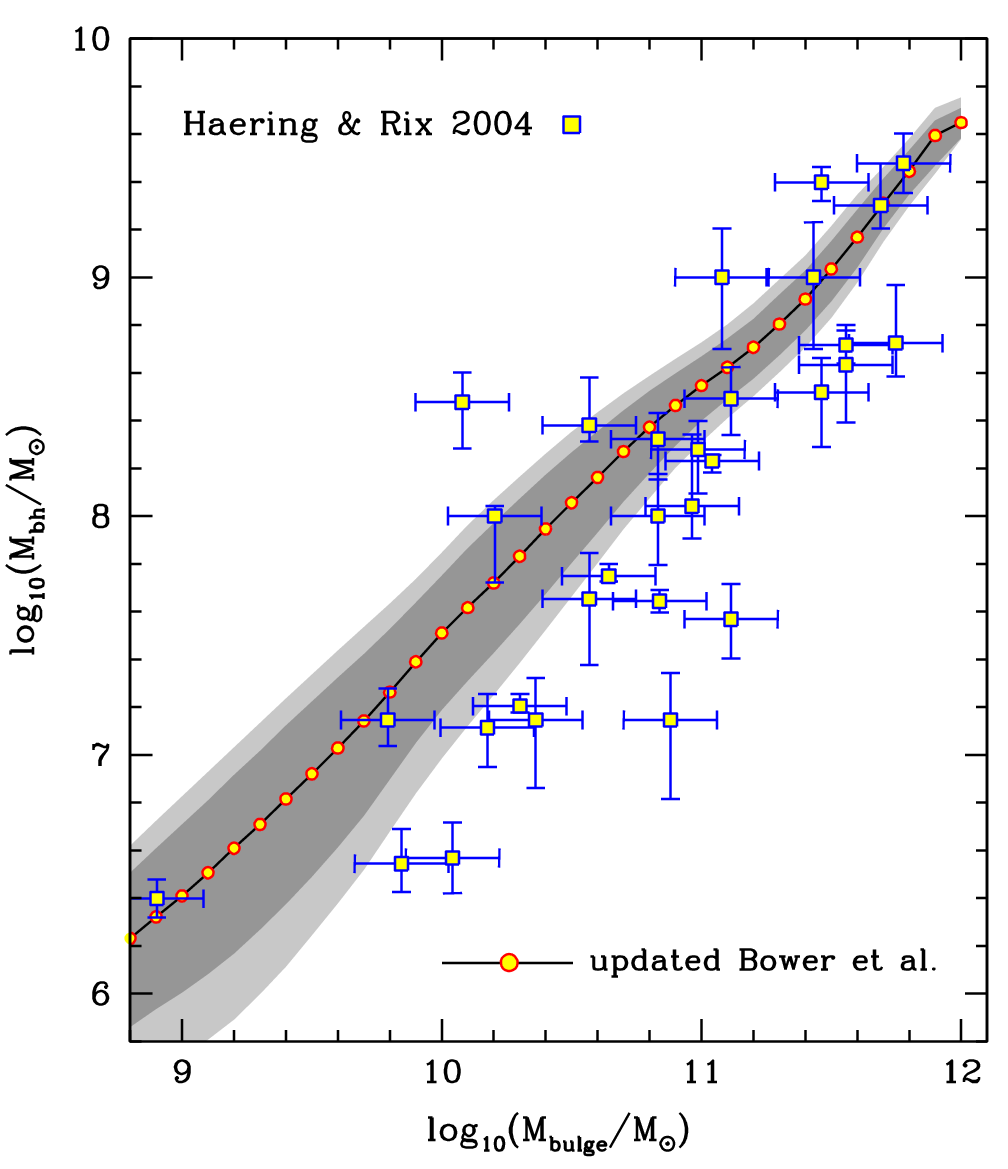}
\caption{The $\MBH-M_{\mathrm{bulge}}$ relation predicted by the updated
Bower \etal (2006) model (solid line). The shaded areas indicate 
the $10-90$ (light) and $20-80$ (dark) percentile spread of the
theoretical predictions. The observational data is taken from H\"aring \&
Rix (2004).}
\label{magorrian}
\end{figure}
\begin{figure}
\center
\includegraphics[scale=0.43]{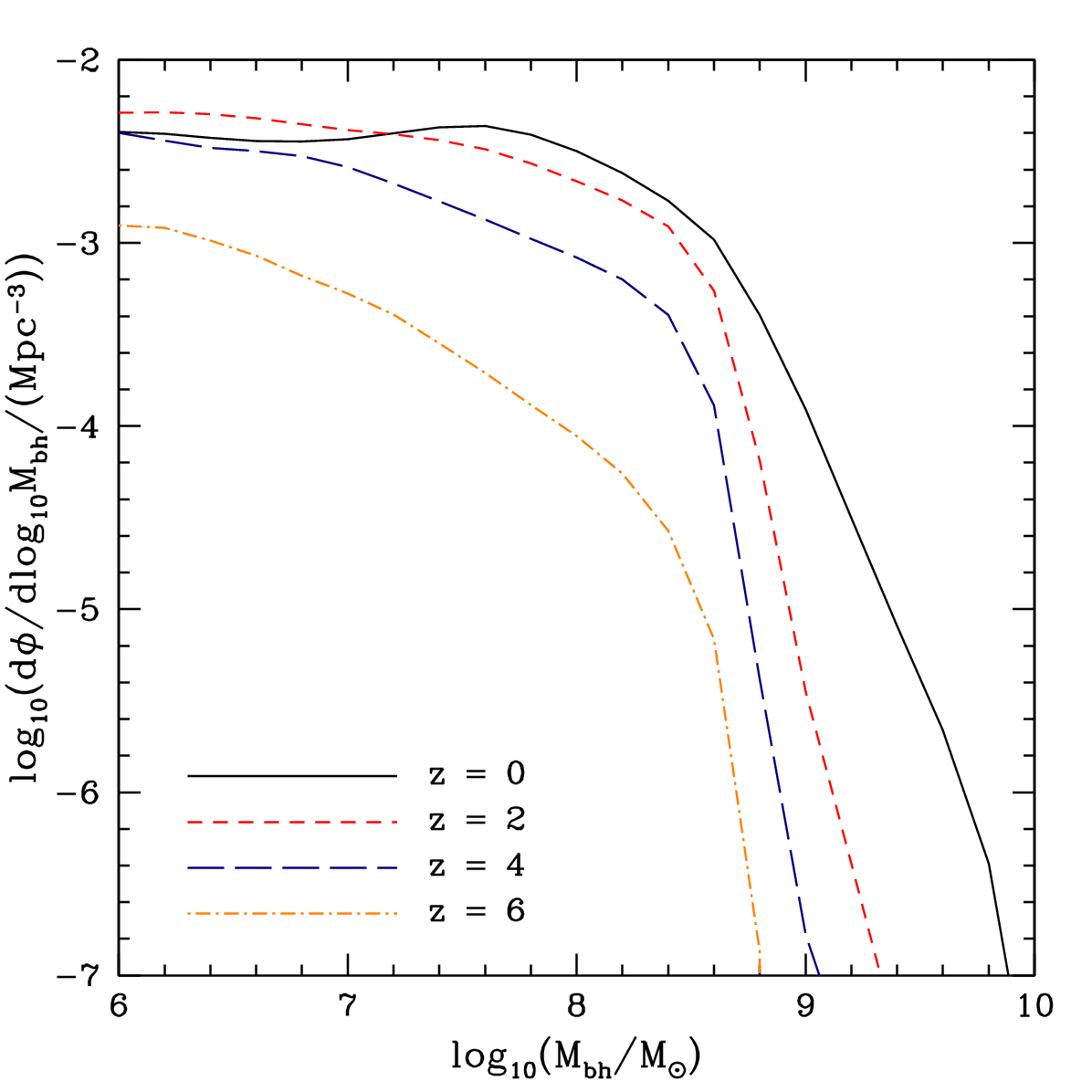}
\caption{The predicted mass functions at $z=0,2,4$ and $6$ for the BH population in our simulation.}
\label{bh_mf}
\end{figure}
By contrast, accretion activity during the radio mode peaks at very
low accretion rates ($\dot{m}\approx10^{-4}$) at $z=0$, with a long
tail to lower values together with some spread to the maximum allowed value of
0.01 (imposed by the AGN feedback model). This upper limit to the mass
accretion rate limits the shift to higher $\dot{m}$ at higher $z$, so
the main evolutionary trend is that there are fewer very low $\dot{m}$ radio
mode objects at high $z$. 

This model results in a {\em bimodal} mass accretion rate distribution for
all BH masses and redshifts. Activity triggered by accretion of cold
gas from the galaxy disc, replenished by galaxy mergers, has a mean
mass accretion rate which is $\sim 1000$ times higher than that of
activity fed by radio mode accretion. However, the relative importance
of these two modes varies with BH mass. This is shown in more detail
in Fig.~\ref{mbh_mdot} which shows the $20-80$ and $5-95$
percentile mass accretion rates as a function of BH mass at $z=0$. More than 5
per cent of the lower mass AGN have accretion via a thin disc
($\dot{m}>0.01$). However, this fraction drops with mass 
at BH masses of $\sim 10^8~\Msun$. More massive 
galaxies are hosted by large ellipticals which are gas poor and are 
dominated by radio mode accretion. 

Fig.~\ref{growth_channels} shows the
contribution from the different growth channels to the final BH mass
at different redshifts. At high redshifts, SMBHs with masses up to
$\sim10^8~\Msun$ build their mass almost exclusively through the 
accretion of cold gas (quasar mode). This forms either a thin disc or an ADAF
depending on $\dot{m}$ (see also Fig.~\ref{mdot}).
Accretion of gas from the hot halo of massive galaxies during the
radio mode always occurs with $\dot{m} \le 0.01$ by construction, 
thus giving rise to an ADAF. This
low upper limit to the mass accretion rate in this mode means that it 
makes comparatively little impact on BH growth, except for masses 
greater than $10^{8}~\Msun$. Note that in the original Bower 
\etal (2006) model, the radio mode is responsible for building most of the 
BH mass above $10^8~\Msun$, while in the updated version presented 
here, both BH mergers and radio mode contribute significantly for high
mass BHs.

In Fig.~\ref{magorrian} we compare the relationship between the BH
mass and host bulge mass predicted by our model to the observational
data. The model predicts an almost linear relation and 
reproduces the observations reasonably well. The scatter 
in BH mass is $\sim1$~dex for bulge 
masses below $10^{11}~\Msun$ and becomes gradually smaller for 
higher bulge masses. The resulting BH mass functions are shown in Fig.~
\ref{bh_mf}. These show that the present day Universe is dominated by 
BHs with masses of $\sim10^{7}-10^{8}~\Msun$ which is in good 
agreement with the recent results from the Sloan Digital Sky Survey 
(SDSS) of Heckman \etal (2004). The shape of the mass functions are fairly 
similar at all redshifts. The main difference occurs at the high-mass end 
where the BH mass function extends to higher masses at low redshifts. For
$\MBH\lesssim10^{8}~\Msun$ the mass function is almost constant
and flat up to $z=2$. This indicates that these BH are already in
place at $z=2$. Above that mass, there is a sharp decrease in the
space density of BH at $z=2$ and $z=4$. At $z=0$, the mass function
has a shallower slope. The estimated BH mass density at 
$z=0$ is $\rho_{\mathrm{bh}}=3.84\times
10^{5}~\Msun~\mathrm{Mpc^{-3}}$, which is in good agreement with 
the value implied by the X-ray background (Fabian \& Iwasawa 1999).

\subsection{Global BH spin distributions} \label{sec:evolution of mass and spin} 

\begin{figure*} 
\center \includegraphics[scale=0.42]{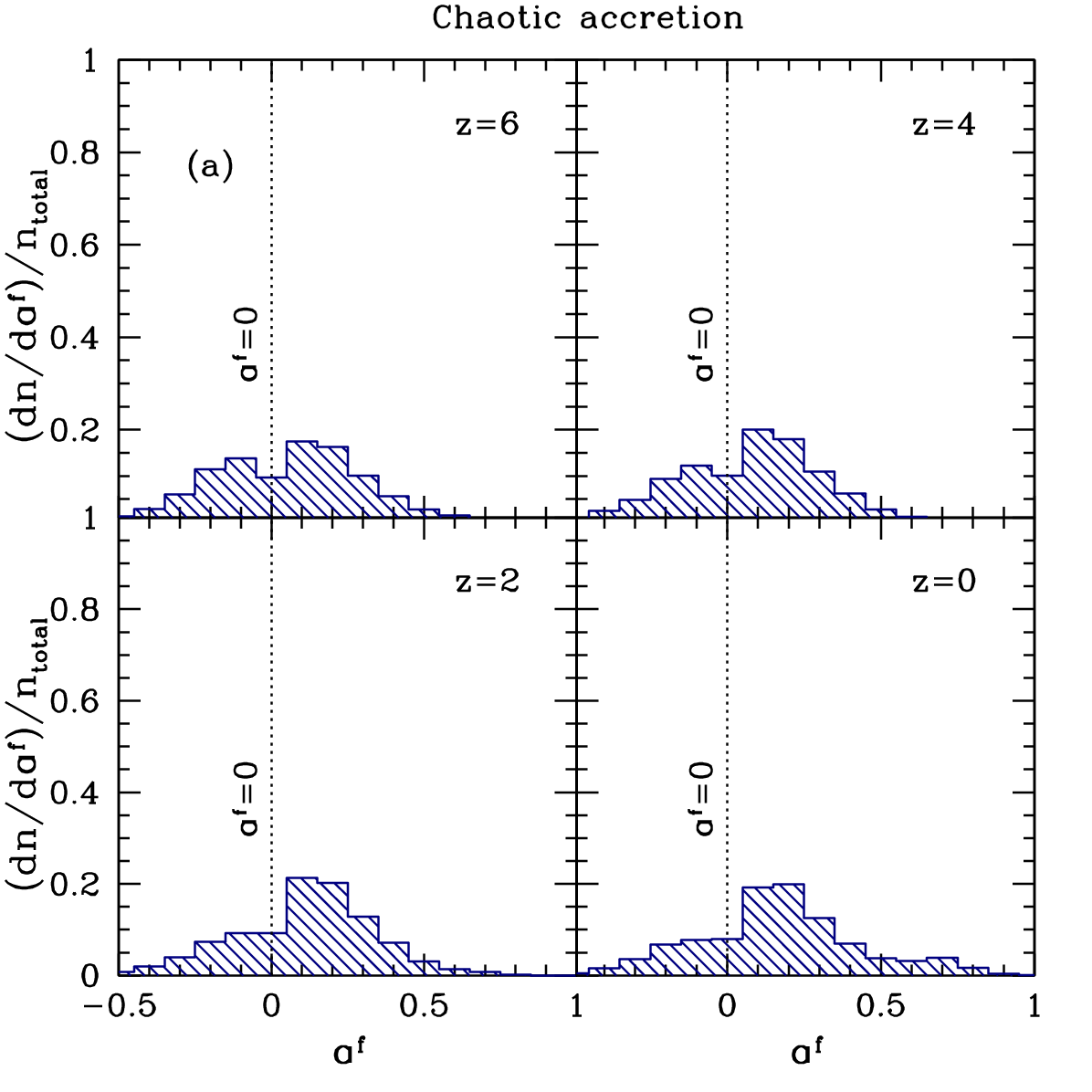} 
\includegraphics[scale=0.42]{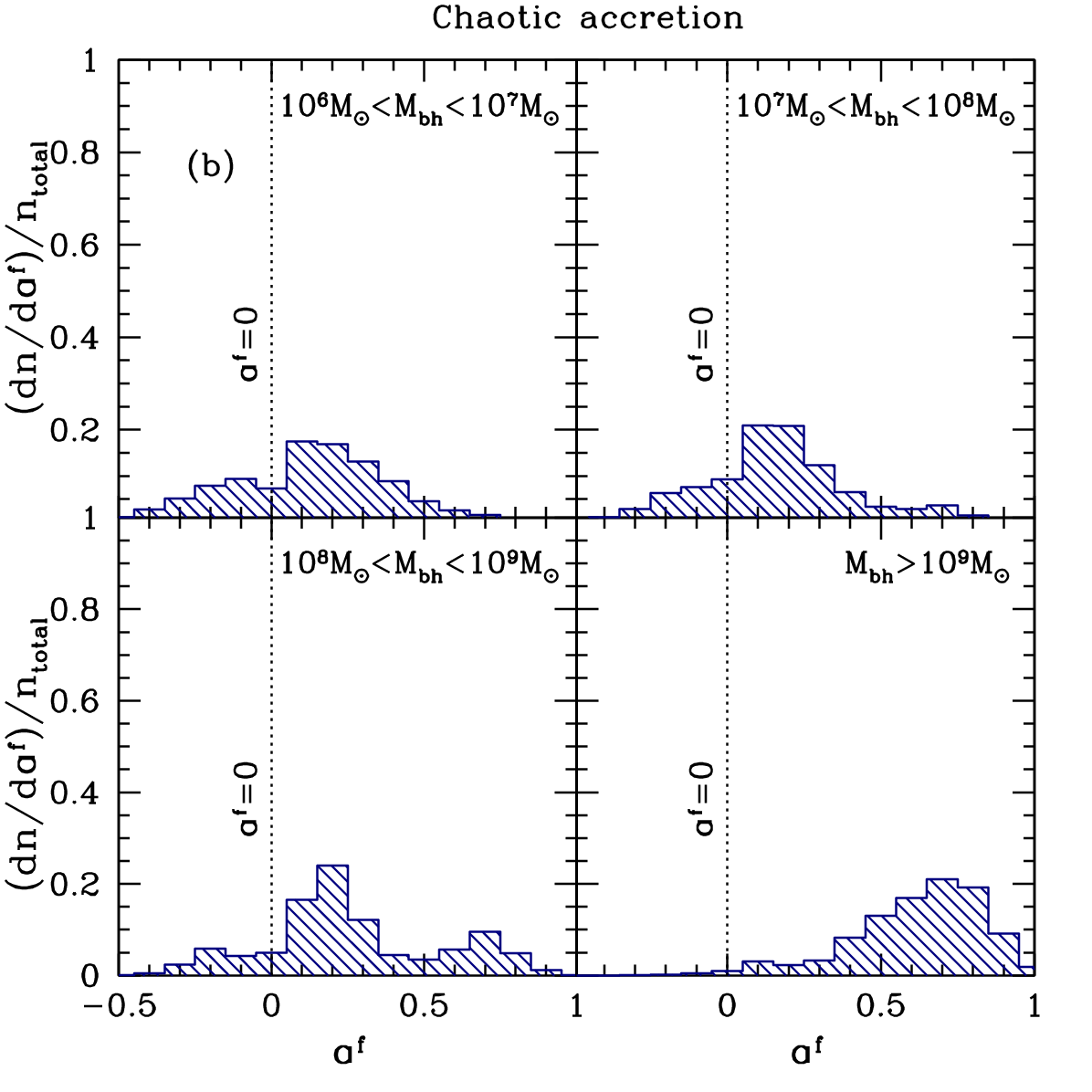} 
\includegraphics[scale=0.42]{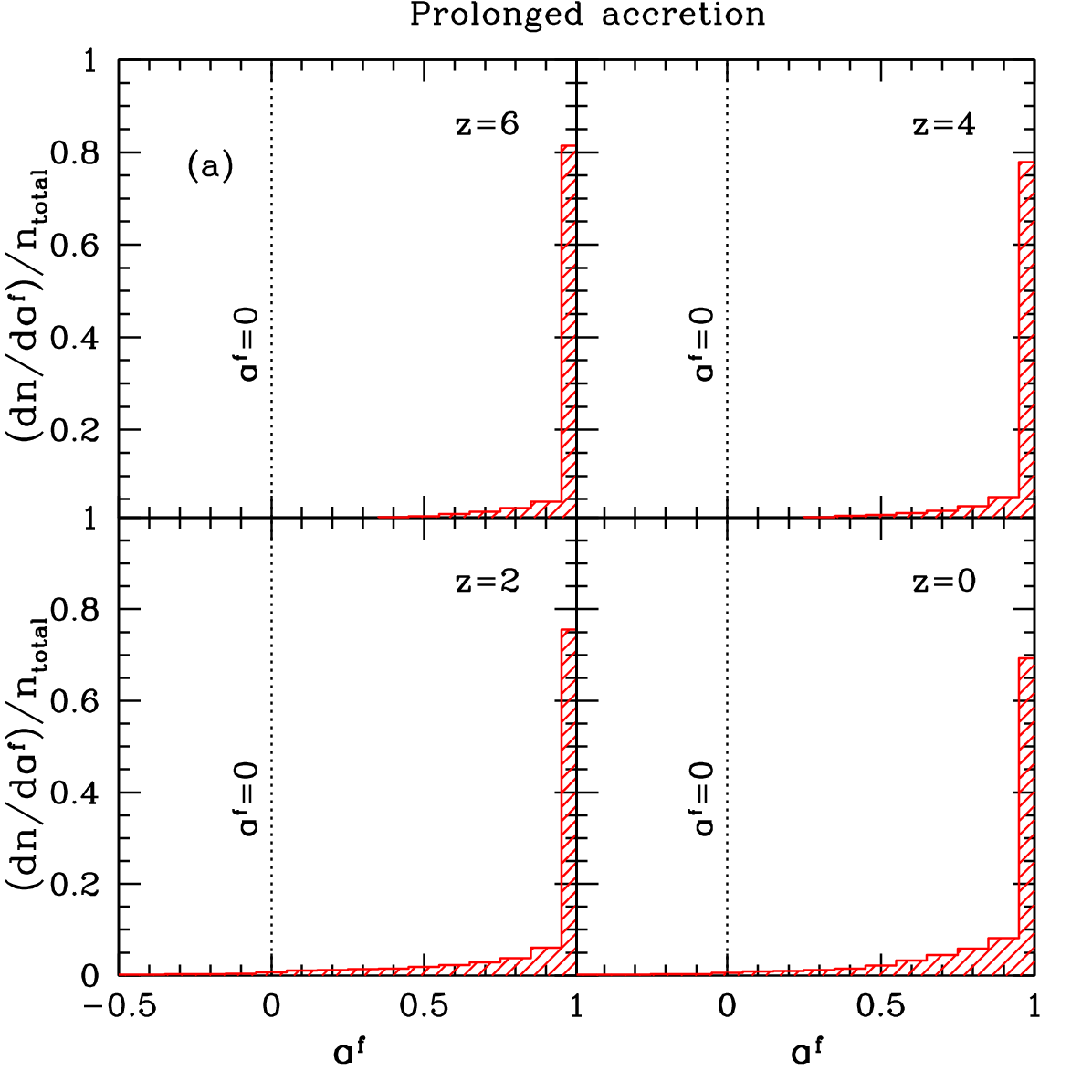} 
\includegraphics[scale=0.42]{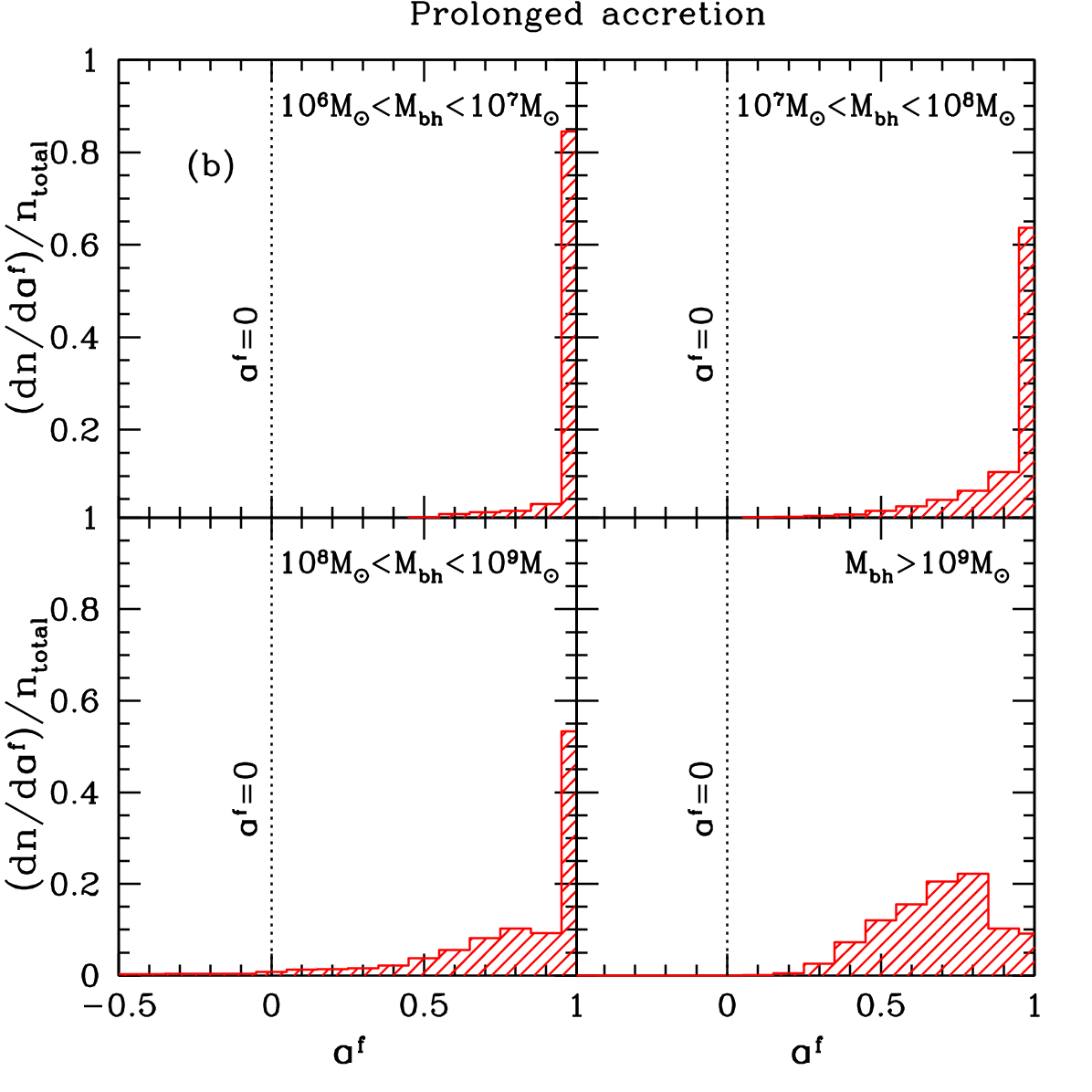} 

\caption{Model predictions for the final BH spin distributions. (a) The spin evolution 
due to BH mergers and accretion and (b) the distribution of BH spins
at $z=0$ in different mass bins in the chaotic (upper panels) and
prolonged (lower panels) accretion models. The main difference
between the two accretion modes is that the chaotic case results in low
mass BHs having spins distributed around zero, whereas the prolonged
case results in a distribution peaked at around $\af=0.998$.}
\label{spin_distributions} \end{figure*}

The histograms in Fig.~\ref{spin_distributions} show the distributions
of BH spins at different redshifts (left panel) and for different mass
ranges (right panel) for all galaxies that host BHs with
$\MBH\geqslant10^{6}~\Msun$. The top and bottom panels correspond
to the chaotic and prolonged accretion models respectively.  
The BH spin distributions predicted by the two accretion 
models are already established at $z=6$ and do not change 
appreciably with time. The major channel for BH growth at $z=6$ is the quasar mode and
this generally takes place via a thin disc. The accreted mass in each
episode is generally larger than the mass of the BH, so this spins it
up to maximal in the prolonged accretion model. The break up of the accretion 
into much smaller individual events in the chaotic
model leads to a random walk spin distribution around a low spin value. 
Nonetheless, the plots in Fig.~\ref{growth_channels} show that BH-BH mergers
contribute to the most massive BHs at late redshifts, as does accretion with
$\dot{m}<0.01$ in both the radio and quasar mode. In order to gain further insight into the effect of these growth channels we show the different spin distributions as a function of BH mass at $z=0$ in the right hand panels.

As already mentioned, low mass BHs are built via
thin-disc accretion in the quasar mode (Fig. \ref{growth_channels}) 
with mean $\dot{m}\sim 0.5$. For a BH of mass $10^6~\Msun$
a typical accretion event in the quasar mode contains $10^6~\Msun$ of
gas, so in the prolonged accretion model this will spin the BH up to
maximal. However, in the chaotic accretion model in which the size 
of the disc is limited by its self-gravity, the mass of the gas
that settles on the disc cannot be more than $\sim (H/R)\times 10^6~\Msun \sim
10^4~\Msun$.  Thus, a $10^6~\Msun$ accretion
event consist of $\sim\negthinspace100$ accretion episodes with
$\Delta M_{\mathrm{episode}}\ll\MBH$, all randomly oriented around the
BH. As a consequence, $10^{6}-10^{8}~\Msun$ BHs experience 
a net spin-down to modest values centred around $a^{f}\sim0.15$.

The growth channel for more massive BHs ($\MBH>10^{8}~\Msun$) includes
a large fraction of objects which accrete in the radio mode or
quasar mode with $\dot{m}<0.01$ i.e. via a geometrically thick
ADAF. However, we do not include full modelling of this process as our spin
evolution calculations always assume a Shakura-Sunyaev thin disc
rather than the appropriate ADAF equations. We do not expect 
this to have a large effect on the spin
distributions as most of the BH mass and spin are built up from high
mass accretion rates.

\begin{figure*}
\center
\includegraphics[scale=0.8]{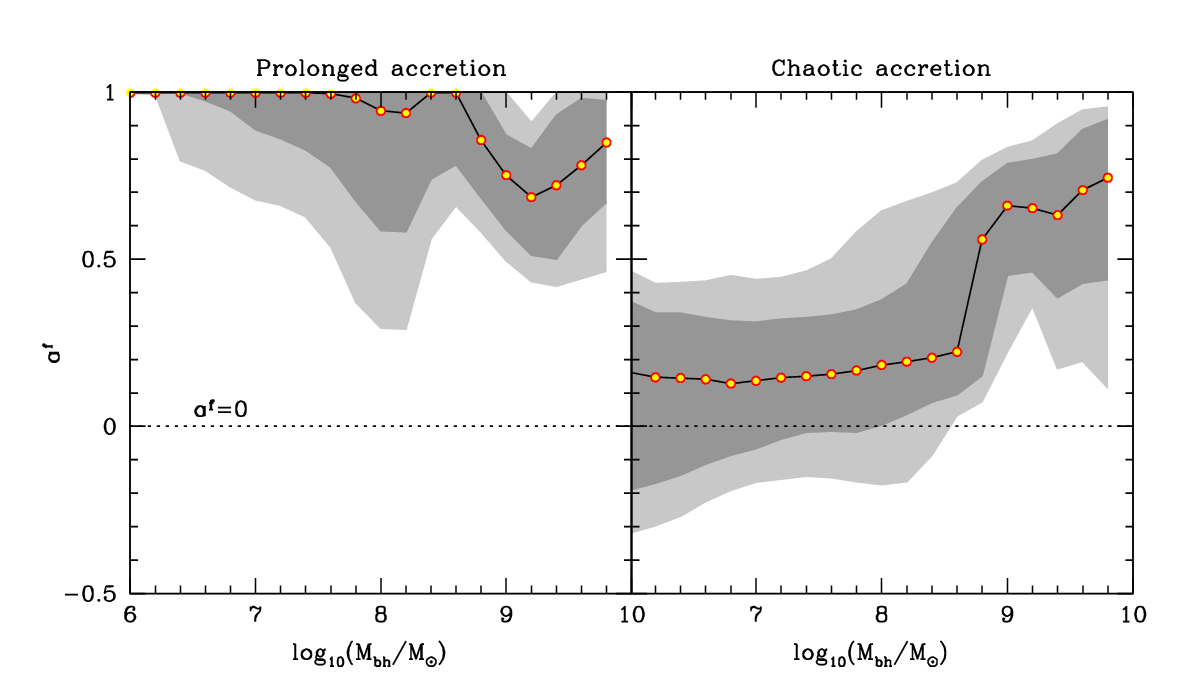}
\caption{The correlation of spin with BH mass (a) for the
prolonged and (b) the chaotic accretion models. The solid lines show
the medians and the shaded areas the $10- 90$ (light) and $20-80$
(dark) percentile spreads of the distributions.}
\label{mbh_spin}
\end{figure*}

The mergers between massive BHs have a significant impact on the final spin distributions. According to the analytic fitting formula, Eqn.~(\ref{rezzolla:final_spin}) from Rezzolla (2008, 2009), the
post-merger spin of the final remnant depends strongly on the masses
of the progenitors. For example, BHs acquiring a final spin greater than $0.69$
after a merger are the end product of binaries of comparable mass ($q\simeq1$)
in which the members already had significant spins (recall that $\af=0.69$ is the final spin of a binary of equal-mass non-spinning BHs). Mergers between BHs of comparable mass are common only for the most  massive BHs ($\MBH\gtrsim10^8~\Msun$) in our simulation\footnote{At redshift zero nearly 19\% of the BHs with $\MBH\geqslant10^6~\Msun$ have experienced at least one merger.
The low merger rate of BHs merely reflects the
relative minor effect that mergers have in the formation of galactic
spheroids (except in very massive ellipticals) in the Bower \etal (2006) model, as found by Parry \etal (2009). For example, at redshift zero
about $6\%$ of the BHs with $\MBH\geqslant10^6~\Msun$ have experienced more
than one merger and only $\sim8\%$ have had a recent major merger}. This is because these BHs are hosted by massive elliptical galaxies that have experienced at least one major merger (merger between galaxies of equal mass) in their past history (Parry \etal 2009). Hence, their BHs, which are of similar mass, will acquire a final
spin close 0.69 or higher (depending on the value of the initial spins) after they coalesce. In contrast to the most massive BHs, lower mass BHs ($\MBH\lesssim10^8~\Msun$) which experience a merger, are involved mostly in minor mergers ($q\ll1$). These do not have a significant impact on the final spin
of the remnant which is dominated by the more massive
member of the binary.

When the effects on BH spin of mergers and accretion are combined, we find the following result. BHs of mass $\gtrsim5\times10^{8}~\Msun$ that have experienced a major merger will acquire spins greater than $\sim0.69$; since these
mergers often occur in gas poor environments, the spin distribution for these masses is not significantly altered by the accretion. For lower mass BHs, accretion \emph{quickly} erases the characteristic post-merger spins, because the growth is dominated by intense accretion during the quasar mode. For chaotic accretion, this results in a bimodal distribution of spins for BHs with mass $10^8-10^{10}~\Msun$. The first peak is located at $\af\sim0.15$ and corresponds to BHs that have had a major merger accompanied by an accretion episode in which the post merger spin is quickly erased and kept low. The second peak is located at $\af\sim0.7-0.8$ and corresponds to BHs that experienced a merger in the absence of accreting flows and acquired a final spin characteristic of post-merger BH remnants.

We note that BHs in our cosmological model have low merger rates at high redshift ($z\sim6$). As redshift decreases, the merger rate increases and reaches a maximum at $z\sim1.5$. Eventually, at $z=0$ nearly 19\% of the BHs with $\MBH\geqslant10^{6}~\Msun$ have experienced at least one merger. Therefore, the significance of mergers increases with decreasing redshift. This accounts for the decrease with redshift of the rapidly rotating BHs in the prolonged model. 

We finally show in Fig.~\ref{mbh_spin} the correlation between the
BH mass and spin. This figure confirms the conclusion of Fig.~\ref{spin_distributions}:
BHs that grow their mass through chaotic accretion
display a strong correlation between their mass and
spin. In this case we find that the BH mass correlates with the host-galaxy
morphology such that small BHs are usually
found in spiral galaxies, whereas massive BHs are found in massive
ellipticals. Thus, the apparent correlation between BH mass and spin
indicates a strong correlation between BH spin and host-galaxy
morphology. As a consequence, we expect to find rapidly rotating BHs
in principle at the centres of very massive elliptical galaxies. 
However, these are also the objects which have low mass accretion
rates (see Fig.~\ref{mbh_mdot}), and so predominantly accrete via an ADAF
whereas some fraction of the lower mass, low spin BH can accrete via
a thin disc. By contrast, in the case of prolonged accretion there is no apparent
correlation between the BH mass and spin since most of the BHs
exhibit very high spin independently of mass. Thus,
rapidly rotating BHs are found in most types of galaxy. The
correlation of mass and mass accretion rate (Fig.~\ref{mbh_mdot})
remains the same, so the most massive ellipticals harbour the most
massive BHs and accrete via an ADAF while a small fraction of the lower
mass BHs in spirals can accrete via a thin disc. Note that the correlation between $\MBH$ and spin for BH masses greater than $10^9~\Msun$ is similar to that predicted by the chaotic model since these BHs include more growth through BH merger activity, which is independent of the accretion model.

\section{Optical luminosities of AGN and the quasar luminosity function}\label{sec:qlf}

\begin{figure}
\includegraphics[scale=0.42]{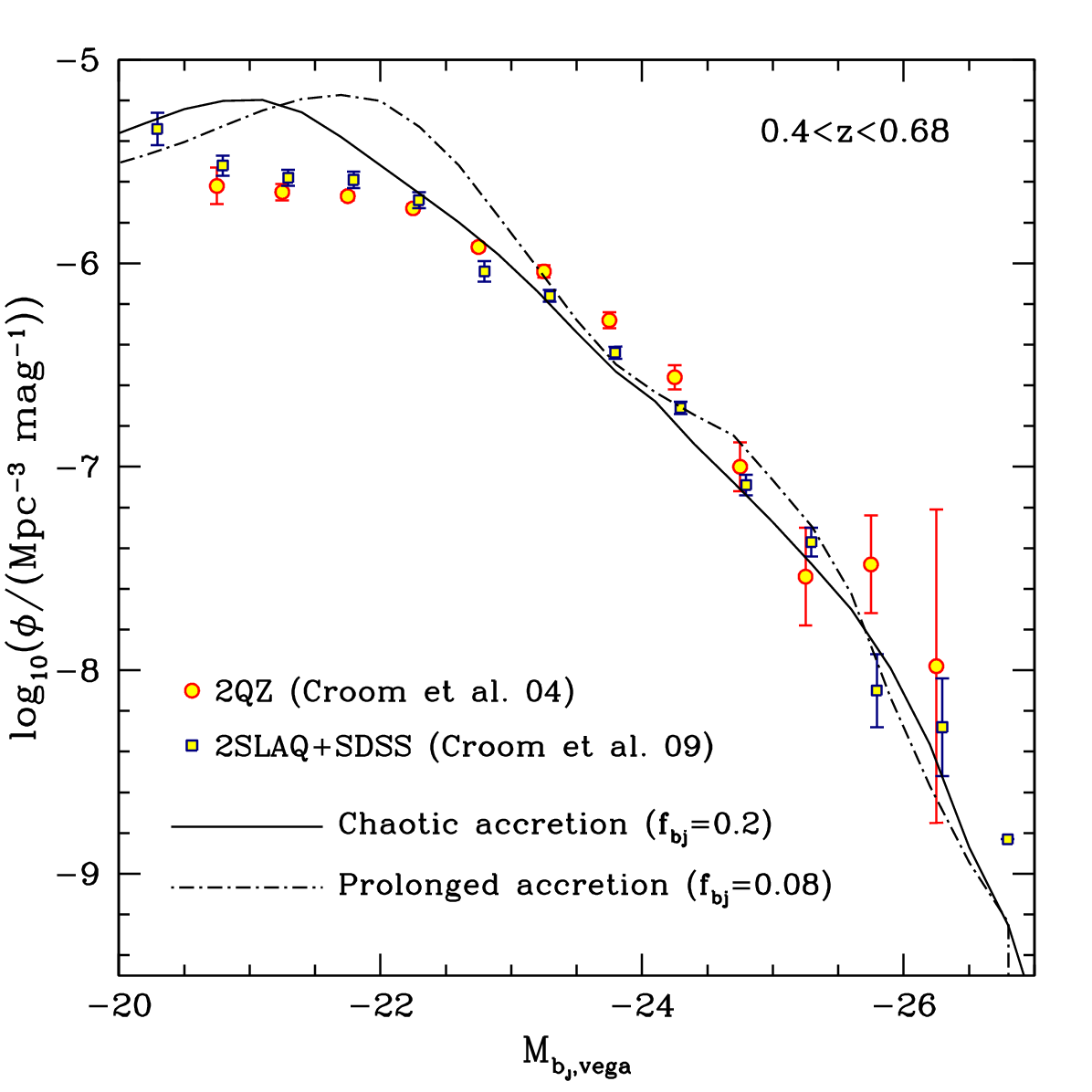}
\caption{The model quasar luminosity function at $z=0.5$ for the
prolonged (dashed-dotted line) and chaotic (solid line) accretion
models. The luminosity function includes only AGN assumed to be
powered by a thin disc ($\dot{m}>0.01$) that accrete in the quasar
mode. The best fits to the observational data require that $8\%$ (prolonged model) and $20\%$ (chaotic model) of the bolometric luminosity be emitted in the $b_\mathrm{J}$-band. It further requires that 40\% of the total number of AGN be obscured in both accretion models. The quasar luminosity functions
from the 2dF (Croom \etal 2004) and 2dF-SDSS QSO survey (Croom \etal 2009) over the range $0.4<z<0.68$ are shown with circles and boxes respectively.}
\label{qlf}
\end{figure}

We now explore whether our models are capable of explaining the observed properties of AGN. In particular, we present our predictions for the quasar luminosity function assuming that quasar activity is driven by accretion onto a SMBH.  In our model, SMBHs 
with $\dot{m}>0.01$ are assumed to accrete
through a thin disc, whereas, those with accretion rates below this
are assumed to accrete through an ADAF.  In the thin-disc
regime the bolometric luminosity is given by the standard expression,
\begin{equation}
L=\epsilon\dot{M}c^2.
\label{disc_bol_td}
\end{equation}  
The accretion efficiency, $\epsilon$, is taken to vary with the hole's spin according to
Eq. (\ref{efficiency}). When the accretion rate becomes super-Eddington 
($\dot{m}>1$), the bolometric luminosity is limited to
$(1+\ln\dot{m})L_{\mathrm{Edd}}$ (Shakura \& Sunyaev 1973). However,
we do not restrict the accretion rate if the flow becomes
super-Eddington. For the ADAFs, we assume that the plasma acquires a two-temperature
configuration. The bolometric luminosity of the disc in this case is
given by Mahadevan (1997),
\begin{eqnarray}
L_{\mathrm{bol,ADAF}}=1.3\times 10^{38}\left(\frac{\MBH}{M_{\odot}}\right)
\left(\frac{\dot{m}^2}{\alpha^2}\right)\left(\frac{\beta}{0.5}\right)\;\mathrm{erg~s^{-1}},
\label{disc_bol_adaf}
\end{eqnarray}
where, $\beta$ is related to the Shakura-Sunyaev viscosity parameter
$\alpha$ through the relation $\alpha\approx0.55(1-\beta)$ (Hawley
\etal 1995). The value of $\alpha$ is taken to be to 0.1 for all objects in our samples.

Observationally, the quasar bolometric luminosity function can be estimated by combining measurements in many different wavelengths (Hopkins \etal 2007; Shankar \etal 2009). We can compare the predictions of our model to the estimated bolometric luminosity function simply by counting the number of quasars radiating at a given luminosity bin. In this analysis, however, we will attempt to calculate the optical $b_{\mathrm{J}}$-band luminosity function since we are ultimately interested in comparing the strength of the radio jet emission of an AGN relative to its disk optical emission (see Section~6).

The conversion factor between bolometric accretion luminosity and $b_{\mathrm{J}}$-band luminosity, $f_{b_{\mathrm{J}}}$, is assumed to be constant and its value is adjusted to match the observed $b_{\mathrm{J}}$-band luminosity function. We take  $f_{b_{\mathrm{J}}}=0.08$ for the prolonged accretion and $f_{b_{\mathrm{J}}}=0.2$ for the chaotic accretion model. The substantial difference between the conversion factors for the two accretion models reflects 
the strong dependence of the bolometric luminosity on the spin through the
accretion efficiency. An increase in spin means that the disc can
extend closer in towards the BH, so the same mass accretion
rate produces more UV emission for the same $b_{\mathrm{J}}$-band
luminosity. Hence the $b_{\mathrm{J}}$-band carries a smaller fraction
of the bolometric luminosity. We check our conversion factors
derived from fitting the quasar luminosity function to those derived
from a Novikov-Thorne (fully relativistic) version of the
Shakura-Sunyaev disc equations, assuming that the $b_{\mathrm{J}}$ band is centred
at $4400$~\AA, with a FWHM of $980$~\AA. 
We find that our assumption is in good agreement with the theoretical calculations as long as we
assume that the disc luminosity is Eddington limited and similarly
they are in good agreement with the values of $0.07-0.15$
suggested by Elvis \etal (1994) from analysis of the SEDs
of a large sample of quasars. Thus, we adopt these
values for the rest of our calculations.

In deriving the luminosity function, we assume that only thin-disc
accreting objects have the intense ionising flux which leads to their
identification as quasars. We further assume that a fraction of $40\%$ of the total number of quasars in obscured in the $b_{\mathrm{J}}$-band by the torus (Polletta \etal 2008). Finally, absolute $b_{\mathrm{J}}$-magnitudes in the Vega
system are obtained using the relation, 
\begin{equation}
M_{b_{\mathrm{J}}}=-10.44-
2.5\log(L_{b_{\mathrm{J}}}/10^{40}\mathrm{erg~s^{-1}}). 
\end{equation}
Fig.~\ref{qlf} shows the resulting quasar luminosity function at
redshift zero for both the prolonged-accretion model, and the chaotic-accretion model.
The models are compared to the observed luminosity functions in the redshift range $0.4<z<0.68$ estimated from the 2dF (Croom \etal 2004) and 2dF-SDSS QSO survey (Croom \etal 2009). The two models agree with each other, although the prolonged-accretion model predicts a somewhat higher volume density of quasars in the $-21\leqslant M_{b_{\mathrm{J}}}\leqslant-23$ and $-24\leqslant M_{b_{\mathrm{J}}}\leqslant-25$ range. The inflection at $M_{b_{\mathrm{J}}}\simeq-24$ that separates the two regions is a reflection of the $10^{8}\Msun$ dip in the $\MBH-a$ correlation seen in Fig.~\ref{mbh_spin}. Both models predict steeper faint end slope ($-20\leqslant M_{b_{\mathrm{J}}}\leqslant-22$) than the observed luminosity functions. A similarly steep faint end is also seen in the soft X-ray luminosity function at $z\sim0.4$ (Miyaji \etal 2001; Hasinger \etal 2005). The differences between the two models in this regime arises because the BHs in the prolonged model that contribute to the faint end have much higher spins, and thus higher bolometric luminosities than in the chaotic model.

In conclusion, we can reproduce the optical luminosity function
reasonably well with either the chaotic or prolonged accretion models.
A consistency check is provided by calculating the average accretion efficiency. Assuming 
only the accreting sources contributing to the quasar luminosity function we find $\langle\epsilon\rangle=14.2\%$ ($\langle a\rangle=0.86$) for the prolonged model and 
$\langle\epsilon\rangle=6.1\%$ ($\langle a\rangle=0.12$) for the chaotic model. The latter is
in good agreement with $\langle\epsilon\rangle=6.7\%$ inferred by optically selected AGN (Mart{\'{\i}}nez-Sansigre \& Taylor 2009) and consistent with the low efficiency that characterises accretion at $z=0$ suggested by Wang \etal (2009).

\section{AGN radio loudness and the spin paradigm} \label{sec:Summary and Discussion} 

In this Section we investigate how the total radio output of an AGN is
related to the accretion process and the central BH properties, and we
present a model where we combine standard accretion disc theory
(including ADAFs), BH evolution and radio jet production to understand
the observed AGN radio loudness.

\subsection{Modelling the jet emission in AGN}

We construct a model for studying the radio loudness of AGN based on
our current understanding of the formation and acceleration of
jets. The starting point is the galaxy formation model described in Section \ref{sec:Cosmological model}.  Galaxies become
active every time the central SMBH experiences an accretion episode,
allowing us to track the evolution of BH mass, spin and mass accretion
rate distributions as described in the previous section. We combine
this with the prescriptions for jet luminosity in the BZ model (Meier
2002). For consistency, we also investigate the more general case where the jet taps energy from both the BH and the accretion disc (Blanford \& Payne 1982), which is described in the hybrid model presented by Meier (2001). 

\subsubsection{The BZ jet model}

AGN jets in the BZ jet model are exclusively powered by
extraction of the rotational energy of the BH and, thus, they are
formed only in AGN hosting rotating SMBHs. The
mechanical energy of a jet is proportional to the square of the
poloidal magnetic field, $B_{\mathrm{pol}}$, at the horizon of the BH  
(Blandford \& Znajek 1977), 
\begin{equation} L_{\rm
{jet}}\propto B_{\mathrm{pol}}^2\MBH^2a^2.  
\label{bz_solution}
\end{equation} 
This is a second-order perturbative solution for the spin parameter $a$ and is used to approximate
the solution for slowly rotating BHs, $a\ll0.998$. By including higher order 
corrections it can be shown that the dependance of $L_{\rm{jet}}$ on $a$ can 
be much steeper for rapidly rotating BHs ($L_{\rm{jet}}\sim a^4$;  Tanabe \& 
Nagataki 2008; see also Tchekhovskoy \etal 2010 for a recent study of
the dynamical range of $L_{\rm{jet}}$ predicted by these solutions).
For this analysis we will adopt the original approximation of Blandford \&
Znajek (1977) for simplicity.

There have been several models that give the strength of $B_{\mathrm{pol}}$ around a
rotating BH.  In most cases, $B_{\mathrm{pol}}$ is expressed in terms
of the azimuthal component, $B_{\phi}$, as $B_{\mathrm{pol}}\approx(H/R)
B_{\phi}$. Therefore, the strength of the jet depends critically on
whether the disc is geometrically thin or thick. For ADAFs ($H\sim R$)
and thin discs (TD; $H\ll R$), this yields (see Meier 2002 and references
therein):
\begin{eqnarray}
L_{\mathrm{jet,ADAF}}=2\times10^{45}\left(\frac{\MBH}{10^9M_{\odot}}\right)\left(\frac{\dot{m}}{0.01}\right)a^2 ~\mathrm{erg~s^{-1}},
\label{adaf_jet_luminosity}
\end{eqnarray}
\begin{eqnarray}
L_{\mathrm{jet,TD}}=2.5\times10^{43}\left(\frac{\MBH}{10^9M_{\odot}}\right)^{1.1}\negthinspace
\left(\frac{\dot{m}}{0.01}\right)^{1.2}a^2 ~\mathrm{erg~s^{-1}}.
\label{td_jet_luminosity}
\end{eqnarray}

In the super-Eddington regime ($\dot{m}\ge 1$), we assume that the
flow remains in a thin disc state as there are as yet no models to describe the
behaviour of the radio jet in this regime. The value of the viscosity
parameter, $\alpha$, in the ADAF and thin-disc regimes is set to
$0.1$. Note that, according to Eqn.~(\ref{adaf_jet_luminosity}) 
the mechanical jet power at the top of the ADAF branch is 
$L_{\mathrm{jet}}=a^2 L_{\mathrm{bol}} \sim 0.01 a^2
L_{\mathrm{Edd}}$~ergs s$^{-1}$. This upper limit to the 
jet luminosity in the ADAF regime is 
the origin of our revision to the fraction of the Eddington
luminosity available for jet feedback into the halo (see Section 2).
For low spin BHs, the BZ jet power drops substantially below this.

\subsubsection{The hybrid jet model}

In the BZ mechanism, the jet is assumed to be powered directly by the
extraction of the rotational energy of the BH, neglecting any energy
that could be extracted from the disc. However, the field lines frozen
to the accreting matter on the disc may generate collimated outflows
as a response to the differential rotation of the plasma, even in the
case when the BH is not rotating (Blandford \& Payne 1982). The
outflow in this case is powered by the extraction of the rotational
energy of the disc, rather than the BH. Thus, in the general case,
both the disc and the hole could contribute to the production of
jets.

Based on these considerations, Meier (2001) proposed a hybrid
model for the radio loudness of AGN in which the jet
luminosity depends on the rotation of both the BH and the accretion
disc. The model includes two distinct accretion states, the ADAF and
thin disc, in which the BH can be rotating or non-rotating. The hybrid
model has a weaker dependence on the spin of the BH than the
standard BZ mechanism. When $a\neq0$ the jet luminosity scales with
spin as
\begin{equation}
L_{\mathrm{jet,ADAF}}^{\mathrm Kerr}\propto(0.55f^2+1.5fa+a^2),
\end{equation}
\begin{equation}
L_{\mathrm{jet,TD}}^{\mathrm Kerr}\propto(1+1.1a+0.29a^2).
\end{equation}
where $f$ and $g$ are parameter related to the angular velocity of the disc and the azimuthal magnetic field respectively (see Meier 2001 for more details; also Nemmen \etal 2007 for a fully relativistic account of the hybrid model). The model appears to be a more realistic interpretation of the jets seen from MHD simulations, since these are still able to power
jets even from Schwarzschild BH accretion. We have tested this model
by coupling it to our predicted spin distributions. However, we find that the
weak dependence on spin in this model results in small
dynamic range in radio luminosity, and the results are very similar to the
predictions of the BZ jet model for the prolonged accretion case (see Section \ref{sec:optical_radio_plane}). We therefore prefer to contrast the prolonged accretion model with the chaotic accretion model coupled to
the spin dependence of the BZ jet model.

\subsection{From jet power to radio luminosity}
\label{From jet power to radio luminosity}
The mechanical BZ jet luminosities presented in the previous section
are converted into radio luminosities according to the theoretical
calculations of Heinz \& Sunyaev (2003) (see also Falcke \etal 1995). Heinz \& Sunyaev derived the non-linear dependence between the jet flux and the physical parameters
$\MBH$ and $\dot{m}$ for radiatively efficient and inefficient
flows. These authors found that the core flux at frequency,
$F_{\nu}$, of a jet scales as $\MBH^{\xi_{1}}\dot{m}^{\xi_{2}}$, with
$\xi_1=\xi_2=17/12$ for ADAF systems and $\xi_1=17/12$, $\xi_2=0$ when
the disc is radiation-pressure supported\footnote{The values for the
scaling indices $\xi_1$ and $\xi_2$ are calculated for a flat
spectrum.}. A similar dependence on the accretion rate and BH mass
seems to be implied by the fundamental plane of BH activity in the
case of radiatively inefficient discs (Merloni \etal 2003; Falcke \etal 2004). Hence,
when jets are launched in ADAFs we assume that
\begin{equation}
L_{\mathrm{R,ADAF}}\propto(\MBH\dot{m})^{1.42}. 
\label{radio_adaf}
\end{equation}
When the ADAF collapses to a thin disc we assume that for the range of
accretion rates we are interested in,
$0.01\leqslant\dot{m}\lesssim100$, the flow is radiation-pressure
supported and thus,
\begin{equation}
L_{\mathrm{R,TD}}\propto\MBH^{1.42} 
\label{radio_td}
\end{equation}

Since we know how $L_{\mathrm{jet}}$ depends on $\MBH$ and $\dot{m}$
we can work out the relation between $L_{\mathrm{R}}$ and
$L_{\mathrm{jet}}$ for both accretion regimes. For example, in the BZ
model, Eqns. (\ref{adaf_jet_luminosity}) and (\ref{td_jet_luminosity})
in combination with Eqns. (\ref{radio_adaf}) and (\ref{radio_td}) give
\begin{equation}
 L_{\mathrm{R,ADAF}}=A_1(\MBH\dot{m})^{0.42}L_{\mathrm{jet,ADAF}},
\label{radio_output_adaf}
\end{equation}
\begin{equation}
L_{\mathrm{R,TD}}=A_2\MBH^{0.32}\dot{m}^{-1.2}L_{\mathrm{jet,TD}}.
\label{radio_output_td}
\end{equation}  
where $A_1$ and $A_2$ are normalisation constants. In order to
constrain the values of these constants, we assume that $A_1/A_2\simeq100$
which arises from the fact that the mechanical power of a jet in an
ADAF is approximately $100$ times higher than that of a jet in a thin disc (Eqns. \ref{adaf_jet_luminosity} \& \ref{td_jet_luminosity}). $A_1$
is then constrained by fitting the predictions of each accretion model for the bright end of the radio luminosity function to the observations (see Section
\ref{sec:rlf}). When we consider the prolonged accretion model this gives 
$A_1=0.05$ and $A_2=A_1/100=0.0005$. For the chaotic 
accretion model, $A_1=0.07$ and $A_2=A_1/100=0.0007$. 
The uncertainty in the parameters $A_1$ and $A_2$ introduces 
an arbitrariness into the model, which, however, is unavoidable since the normalisation of $L_{\mathrm R}$ is not calculable from first principles.

\begin{figure}
\center
\includegraphics[scale=0.43]{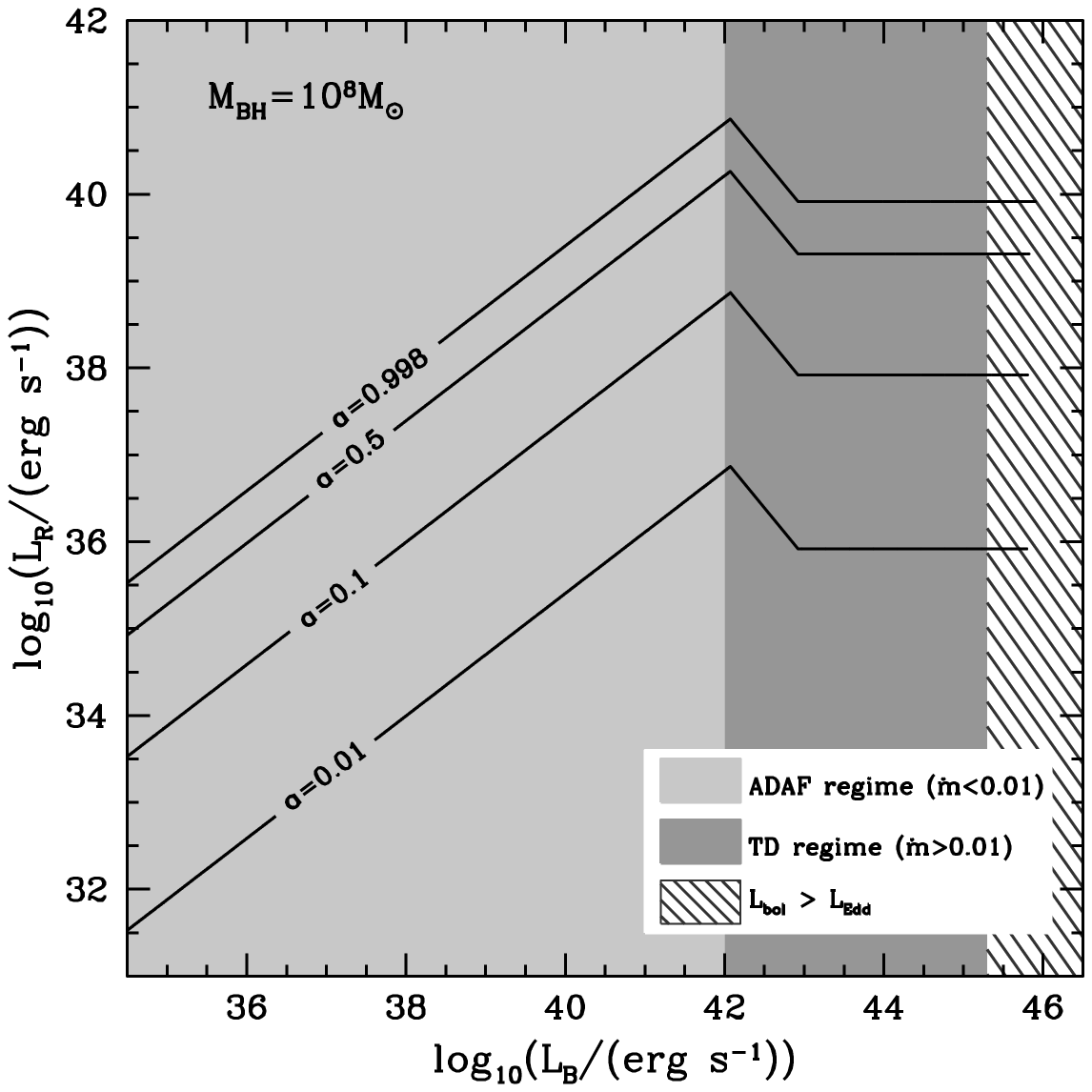}
\caption{The predictions of the BZ model for the optical and radio
luminosities of an AGN. Results are shown for an AGN hosting a SMBH
with $\MBH=10^8~\Msun$ and rotating with spin $a=0.01, 0.1, 0.5$ and
$0.998$. The viscosity parameter $\alpha$ is fixed to 0.1 in both
the ADAF and thin-disc regimes. We represent the different accretion
regimes with different background shadings. The sharp drop in radio
luminosity seen in all cases is due to the transition form the ADAF
to the thin disc regimes when the accretion rate exceeds
$0.01L_{\mathrm{Edd}}$. The rightmost shaded area depicts the region
where accretion becomes super-Eddington. In this region, the optical
luminosity is set to $(1+\ln\dot{m})L_{\mathrm{Edd}}$. The
discontinuity above $10^{42}~\mathrm{erg~s^{-1}}$ in accretion
luminosity is due to the fact that thin discs are radiatively
efficient and thus, for the same accretion rate, are more luminous
than an ADAF. }
\label{jet_model}
\end{figure}

\subsection{The $\log L_{\mathrm{B}}-\log L_{\mathrm{R}}$ plane for a $10^{8}~\Msun$ BH}
 
The theoretical predictions of the BZ model for the
evolution of AGN jets on the $\log L_{\mathrm{B}}-\log L_{\mathrm{R}}$
plane are illustrated in Fig.~\ref{jet_model}. The optical luminosity,
$L_{\mathrm{B}}$, is obtained as discussed in Section 4.
The jet strength is explicitly determined by the three fundamental parameters of our
model, the spin, the accretion rate and the BH mass (and also by the
values of the normalisation constants $A_1$ and $A_2$). The
predictions of the models are shown for a BH mass of $10^{8}~\Msun$
spinning at 
$a=0.01, 0.1, 0.5$ and $0.998$. The different accretion regimes
are represented by different background shadings. As shown
in the diagram, a BH that accretes through an ADAF develops jets whose
radio luminosity increases with the accretion rate on the
$\log L_{\mathrm{B}}-\log L_{\mathrm{R}}$ plane. When the accretion rate
becomes higher than 0.01, the accretion flow enters the thin-disc
regime and the jet collapses by a factor roughly equal to the
difference in $H/R$ between the two very different accretion flows.

In the thin disc regime, the mechanical jet power increases with
$\dot{m}$ but this does not translate into an increase in observed
radio flux. Instead, the radio flux remains constant, while the
optical flux increases with mass accretion rate. Thus, the much lower
radio emission is sustained throughout the thin disc regime and we do
not expect objects powered by thin discs to be radio
loud. Our model extends this behaviour to super-Eddington mass
accretion rates, although we note that this is almost certainly an
underestimate of the jet and radio luminosities as $H/R$ for such flows
increases again to $\sim 1$. Thus, there is a clear accretion mode
switch in the models. The brightest radio sources are the highest spin
BHs accreting at the maximum rate for an ADAF. Since this is
$L/L_{\mathrm{Edd}}=0.01$, this maximum radio luminosity scales also with
the mass of the BH.

\subsection{The radio luminosity function}\label{sec:rlf}

\begin{figure}
\includegraphics[scale=0.43]{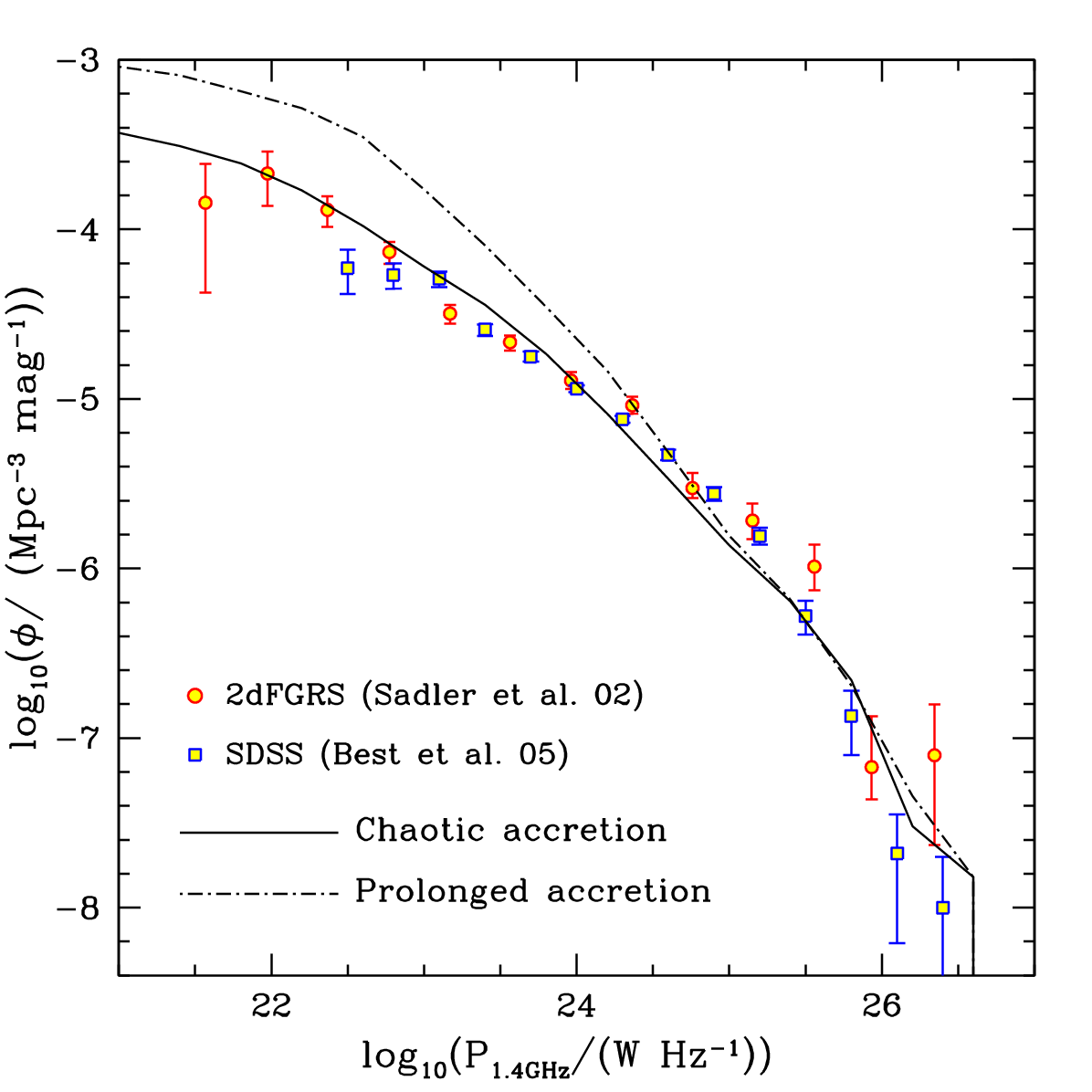}
\caption{The radio luminosity function of AGN at 1.4 GHz. The lines 
show our model predictions for the prolonged (dashed-dotted) and
chaotic (solid) accretion models. The symbols show
observational measurements for the local radio luminosity function
from Sadler \etal (2002) for the 2dFGRS (circles) and Best \etal
(2005) for the SDSS (boxes). The predictions of the BZ model, in
combination with the chaotic accretion model, reproduce the observed
radio luminosity function reasonably well.}
\label{rlf}
\end{figure}

A first test of the jet model is the radio luminosity function of
AGN. In Fig.~\ref{rlf} we show the predictions of the BZ jet model for
the radio luminosity function in both the prolonged and chaotic accretion
cases and compare with the estimates from the
Two-degree-Field Galaxy Redshift Survey (2dFGRS; Sadler \etal 2002) 
and from the SDSS (Best \etal 2005)\footnote{The results from 2dFGRS have been adjusted to
the cosmology adopted by Best \etal (2005).}. The model radio luminosity
function is derived for systems powered by ADAFs and thin discs,
including super-Eddington objects, in a simulation volume of
$V=1.25\times10^{8}~\mathrm{Mpc}^{-3}$. 

The coupling between the different spin distributions and the BZ
mechanism gives interestingly different predictions for the two
accretion models. The chaotic model shows a fairly good agreement with
the observations throughout the entire energy spectrum. The faint end
is dominated mainly by galaxies that host slowly rotating BHs whereas
the bright end is dominated by galaxies that host rapidly rotating massive
BHs. The majority of these BHs accrete in the radio mode. A characteristic break in the slope at
$\sim10^{25}~\mathrm{W~Hz^{-1}}$ separates the two regimes.

In contrast, in the prolonged accretion case the same sample of galaxies
shows a different distribution of luminosities. The model
overproduces the number density of faint radio sources by a factor of
$\sim3$ compared to the chaotic model, due to the fact
that these luminosity bins are populated by galaxies with rapidly
rotating BHs whose radio luminosity is too larger. This trend
extends down to luminosities of $10^{25}~\mathrm{W~Hz^{- 1}}$. Above
that luminosity, the model is in reasonably good agreement with the
observations. Since the mechanism that spins those BHs
up, namely the BH mergers, is present in both models, the two predicted
luminosity functions agree with each other very well at
$10^{25}-10^{26}~\mathrm{W~Hz^{-1}}$.  Note that the parameters $A_1$ and $A_2$ introduced in Section~\ref{From jet power to radio luminosity} affect only the normalisation of the radio luminosity function. The shape is a prediction of the model and depends only on the distribution of BH masses, spins, and accretion rates.

\section{Predictions for the  $L_{\mathrm{B}}-L_{\mathrm{R}}$ AGN activity}
\label{sec:optical_radio_plane}

\begin{figure*}
\center
\includegraphics[scale=0.43]{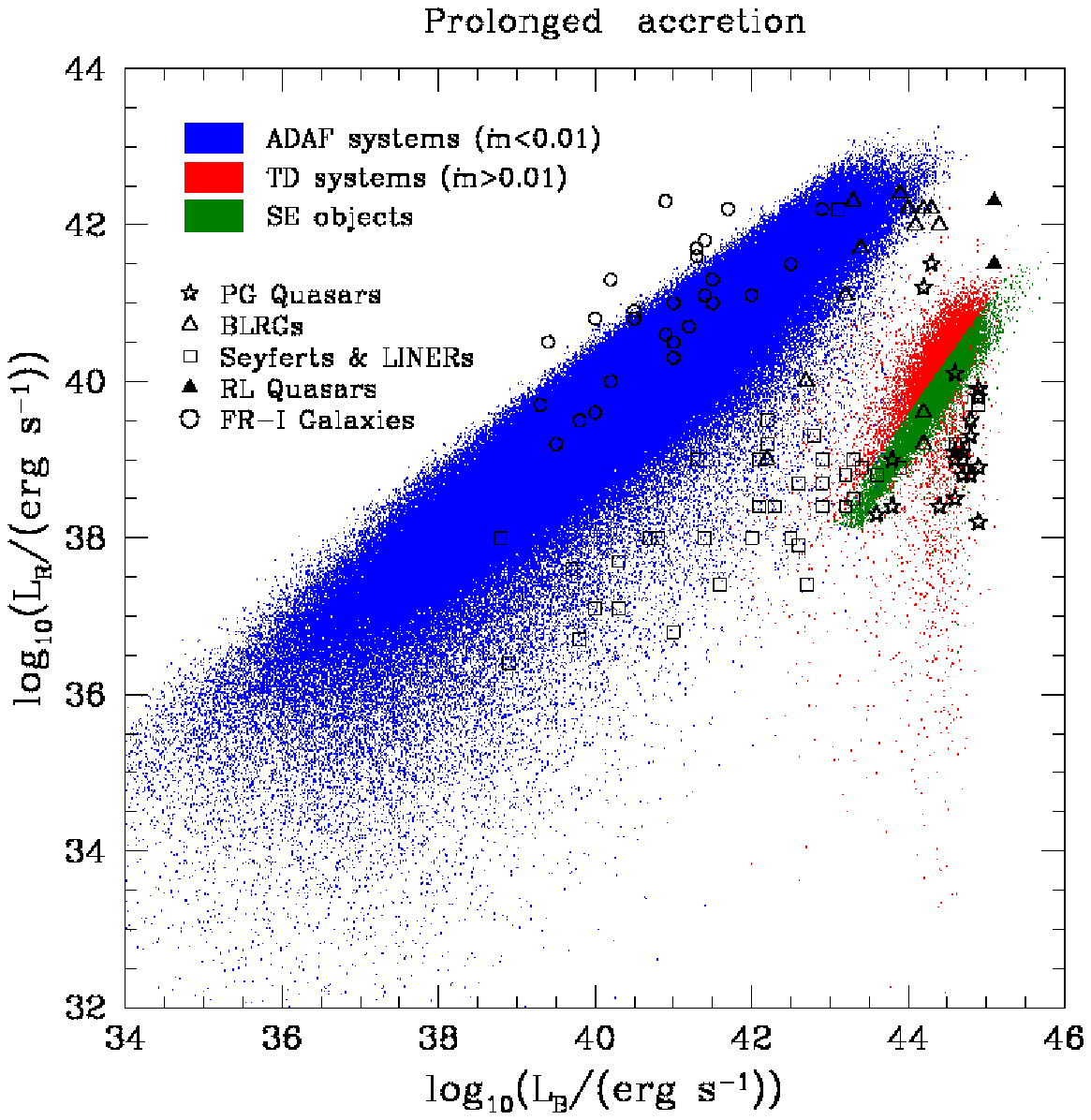}
\includegraphics[scale=0.43]{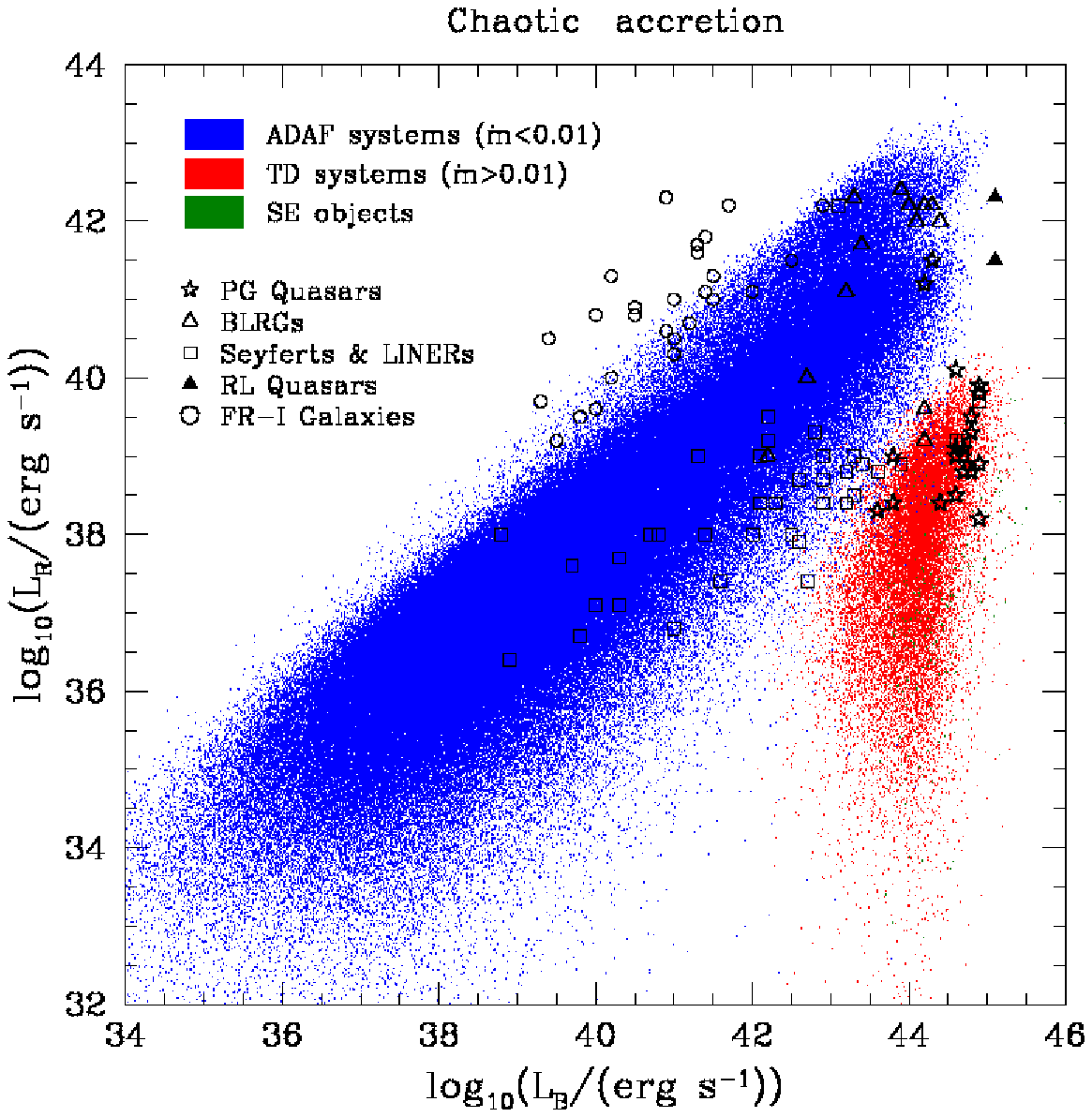}
\caption{Radio luminosity \textit{vs}. $B$-band nuclear luminosity
for the BZ model in (a) the prolonged and (b) the chaotic accretion
models. Only galaxies with $z<0.14$, $M_{\mathrm V}<-20.5$ and 
$M_{BH}>10^6\mathrm{M_{\odot}}$ are shown. The blue points represent
galaxies whose SMBHs accrete via an ADAF. These include galaxies
experiencing \emph{both} the radio and quasar modes. The red points
represent galaxies powered by thin discs around SMBHs. The green
points denote super-Eddington objects. The observational data are
taken from Sikora \etal (2007): BLRGs are shown by open triangles; 
radio-loud quasars by filled triangles; Seyfert galaxies and LINERs
by open squares; FRI radio galaxies by open circles; and PG Quasars
by open stars.}
\label{jet_luminosities}
\end{figure*}

\subsection{The distribution of galaxies on the optical--radio plane}

We now explore the distribution of AGN on the $\log
L_{\mathrm{B}}-\log L_{\mathrm{R}}$ plane. We consider a sample of 
galaxies brighter than $L_{\star}$ ($M_{\mathrm V}\simeq-20.5$). 
This distribution is displayed in Fig.~\ref{jet_luminosities}, 
where we plot the $B$-band versus the
radio luminosity of AGN for both the prolonged and
chaotic accretion models (left and right plots respectively). All galaxies are volume
weighted and those accreting through a thin disc are weighted according to 
the fraction of their lifetime during which they are active. The
colour coding corresponds to different types of accretion: blue
shows AGN powered by an ADAF, red those powered by a thin disc and
green indicates the super-Eddington objects.

Along with the model predictions, we plot the radio and optical
luminosities of a sample of AGN-powered radio sources studied by Sikora
\etal (2007). We include only local objects i.e. those at
$z<0.14$, so this excludes the very brightest radio and optical
sources. The evolution of the sources across cosmic time will be explored in a future paper. 

The data form two distinct sequences on the $\log
L_{\mathrm{B}}-\log L_{\mathrm{R}}$ plane. The upper sequence
represents radio-selected galaxies and the lower sequence
optically-selected objects. The objects in the upper sequence are on
average $\sim\negthinspace3$ orders of magnitudes more radio loud than
those in the lower sequence. These radio bright objects are Fanaroff and Riley
class I (FR-I) galaxies, broad line region galaxies (BLRGs) and radio
loud quasars (RLQs). These are thought to be powered by very
massive BHs (Mclure \& Dunlop 2002), hosted by elliptical
galaxies. Objects in the lower, radio-quiet sequence include
Seyferts, low ionisation emission regions (LINERs) and radio-quiet
quasars (RQQs). 

The observational data suffer from multiple selection effects, which
may be responsible for the apparent dichotomy seen in the distribution
(see discussion by Sikora \etal 2007). The fact
that some FRI galaxies have much higher radio power than predicted by
our model could be due to the fact that these data include {\em lobe}
power, whereas we only model the {\em core} radio emission. However, the
locus of the data points sets upper limits for the radio luminosity
of radio-quiet and radio-loud objects, and provides a rough visual
description of the general bulk properties of the different populations.

In both accretion models, AGN span a wide range of optical and
radio luminosities. AGN powered by ADAFs generically produce more
radio and less optical emission than those powered by a thin
disc. Thus, the accretion mode switch at $\dot{m}=0.01$ from ADAF to
thin disc produces a marked transition in radio loudness.  All our
models show that the radio-loud sequence lies at the upper end of the
radio luminosity envelope for SMBH accreting via a hot
flow. Conversely, the radio-quiet locus lies predominantly on the
thin disc points, although the lowest luminosity radio-quiet AGN
(LINERS) lie on the lower end of the ADAF regime. Thus, the {\em major}
switch between radio--loud and radio-quiet is the change in jet
properties as the accretion flow collapses from a hot, geometrically
thick configuration, to a cool, geometrically thin disc (Jester 2005).

Nonetheless, there can also be an additional spin dependence which enhances
the difference in radio properties, although this depends strongly
on the details of the jet modelling and the spin distribution. The BZ
jet models are strongly dependent on spin, so their radio luminosity
amplifies differences in the spin distribution.  In the prolonged
accretion model almost all SMBHs have maximal spin, so there is little
dispersion in radio flux. Conversely, for the chaotic model, the
correlation between mass and spin means that the low mass and hence
lower optical luminosity objects have dramatically lower radio power
than the higher mass BH in both the ADAF and thin disc regime.

There are also subtle differences in optical luminosity between the
prolonged and chaotic models, as the higher efficiency of high spin
accretion means that the same mass accretion rate gives rise to a
higher luminosity.Thus, there are more super-Eddington sources 
in the prolonged accretion model, in which low mass objects have 
high spin, than in the chaotic model.

\subsection{The distribution of galaxies on the $\rdl-\lambda$ plane}

In addition to the AGN optical and radio output, the model allows us to study how the \emph{radio loudness} $\rdl$ of AGN depends on various physical parameters such as the BH mass and the Eddington ratio $\lambda=L_{\mathrm{bol}}/L_{\mathrm{Edd}}$. The radio loudness measures the radio to optical flux ratio, $\rdl\equiv L_{\nu_{\mathrm{R}}}/L_{\nu_{\mathrm{opt}}}$ (Richards \etal 2006). Following the definition of $\rdl$ in Sikora \etal (2007), we consider the total radio flux at 5GHz, $L_{R}=\nu_{5_{\mathrm{GHz}}}L_{\nu_{5_{\mathrm{GHz}}}}$, and therefore we express the radio loudness as $\rdl=(\nu_{\mathrm B}/\nu_{5_{\mathrm{GHz}}})\times L_{\mathrm R}/L_{\mathrm B}=1.36\times 10^{5}L_{\mathrm{R}}/L_{\mathrm{B}}$, given that the B-band is centred at a wavelength of $4400$\AA.

\begin{figure}
\center
\includegraphics[scale=0.43]{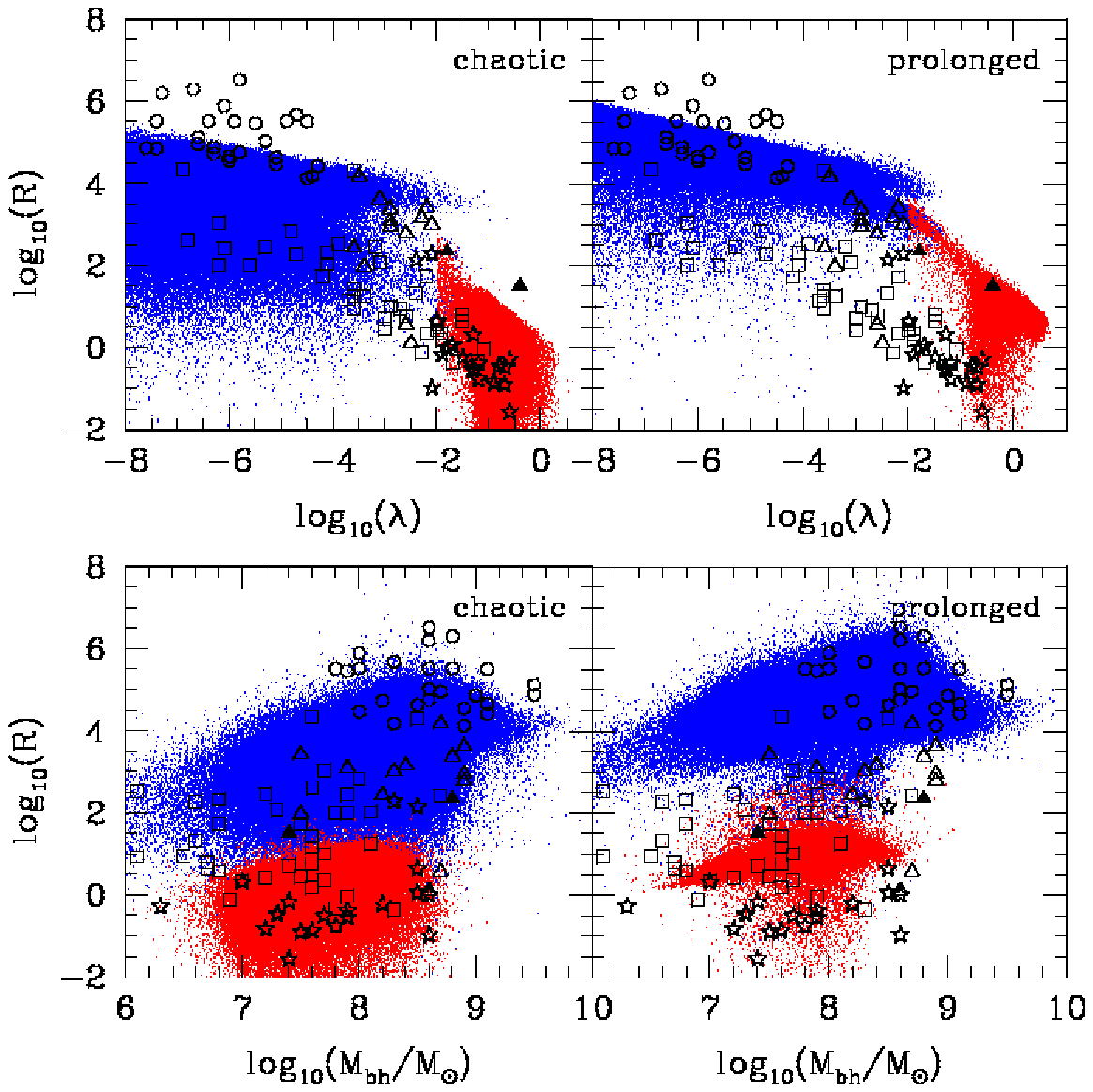}
\caption{Scatter plots of radio loudness ($\rdl=(\nu_{\mathrm B}/\nu_{5_{\mathrm{GHz}}})\times L_{\mathrm R}/L_{\mathrm B}$) \textit{vs.} BH mass (top panels) and $\lambda$ parameter (bottom panels) in the chaotic and prolonged accretion models for the sample in Fig.~\ref{jet_luminosities} (galaxies with $z<0.14$, $M_{\mathrm V}<-20.5$ and
$\MBH>10^6\mathrm{M_{\odot}}$). Different colours represent different accretion regimes: blue for ADAF systems and red for thin-disc systems. The observational data are
taken from Sikora \etal (2007): BLRGs are shown by open triangles, 
radio-loud quasars by filled triangles, Seyfert galaxies and LINERs
by open squares, FRI radio galaxies by open circles, and PG Quasars
by open stars.}
\label{radio_loudness}  
\end{figure}

We plot the theoretical predictions for the radio loudness of the AGN in our sample along with the observational data set of Sikora \etal (2007) in Fig.~\ref{radio_loudness}. Our predictions suggest a clear inverse correlation between $\rdl$ and $\lambda$ with a substantial scatter in both accretion models (upper panels in Fig.~\ref{radio_loudness}). In the ADAF regime, the distribution of objects shows a correlation of the form $\rdl\propto\lambda^{-0.4}$ (one can derive the correlation between $\rdl$ and $\lambda$ using Eqns. (\ref{radio_output_adaf}), (\ref{adaf_jet_luminosity}) and (\ref{disc_bol_adaf})) which is driven mainly by the strong dependence of the disc luminosity on the accretion rate, $L_{\mathrm{bol}}\propto\dot{m}^{2}$. The correlation between $\rdl$ and $\lambda$ becomes steeper in the thin-disc regime, where the data approximately follow $\rdl\propto\lambda^{-1}$. The dependence of $\rdl$ on $\lambda$ in the thin-disc regime arises from $\rdl\propto L_{\mathrm{bol}}^{-1}\propto\dot{m}^{-1}\propto\lambda^{-1}$.

In both models, the radio-quiet AGN ($\log\rdl<1$, following Ho 2002) are preferentially found in the $\log\lambda\gtrsim-2$ regime, while the radio-loud sources ($\log\rdl>$1) populate exclusively the $\log\lambda\lesssim-2$ regime. When compared to the data from Sikora \etal (2007), the predictions of the chaotic model give an adequate representation of the AGN on the $\rdl-\lambda$ plane. The model reproduces well the loci of the different AGN populations and the slope of the overall distribution. Objects such as quasars and Seyferts are radiating at $\sim(0.01-1) L_{\mathrm{Edd}}$, while the radio galaxies are characterised by very low sub-Eddington luminosities ($<0.01 L_{\mathrm{Edd}}$).

By contrast, in the prolonged model there is an absence of objects with $\log \rdl<3$ in the $-4\lesssim\lambda\lesssim3$ regime. This is mainly because most objects pile up in the upper envelope of the distribution due to their high radio luminosities (and BH spin values). We note again that FR-I galaxies (open circles) have higher values of radio loudness than the models. This could be due to the fact that the measured radio luminosities for these objects include power from the lobes. 

The objects in our sample populating the radio-loud part of the $\rdl-\lambda$ plane are associated mainly with giant ellipticals that host very massive BHs (see below). Accretion onto these BHs in the ADAF regime produces significant radio power because the spin is very high. In combination with the low optical luminosities of these objects, this gives rise to very high $\rdl$ values. Therefore, these specific properties of the host of the radio-loud AGN may suggest a correlation between $\rdl$ and $\MBH$. However, when we plot $\rdl$ versus $\MBH$ (lower panels in Fig.~\ref{radio_loudness}) we find that in both models there is no apparent correlation between these two quantities. The data display significant scatter along the $\MBH$ axis which is due to the complex dependence of $\rdl$ on $\MBH$ through the optical and jet luminosity expressions in Eqns. (\ref{radio_output_adaf}), (\ref{radio_output_td}), and (\ref{disc_bol_adaf}). Both models reproduce the loci of the different AGN population in the Sikora \etal data set remarkably well. ADAF systems, mainly identified with radio-loud sources, span a wide range of BH masses ($10^{6}-10^{10}~\Msun$). In contrast, radio-quiet objects have BH masses in the range $\sim10^{7}-3\times10^{8}~\Msun$. Finally, in both models all the sources hosting BH with $\MBH\gtrsim3\times10^{8}~\Msun$ are exclusively radio-loud. 

\subsection{A physical view of the ``chaotic accretion" population}

While both accretion models give a fair representation of the locus of
the heterogeneous radio-optical data points from Sikora \etal (2007),
the chaotic accretion model provides a better fit to the much more homogeneous
data from the radio luminosity function described in Section~\ref{sec:rlf}. Hence, in this section, we focus on the chaotic accretion model and explore the
results in greater detail.

\begin{figure}
\center
\includegraphics[scale=0.365]{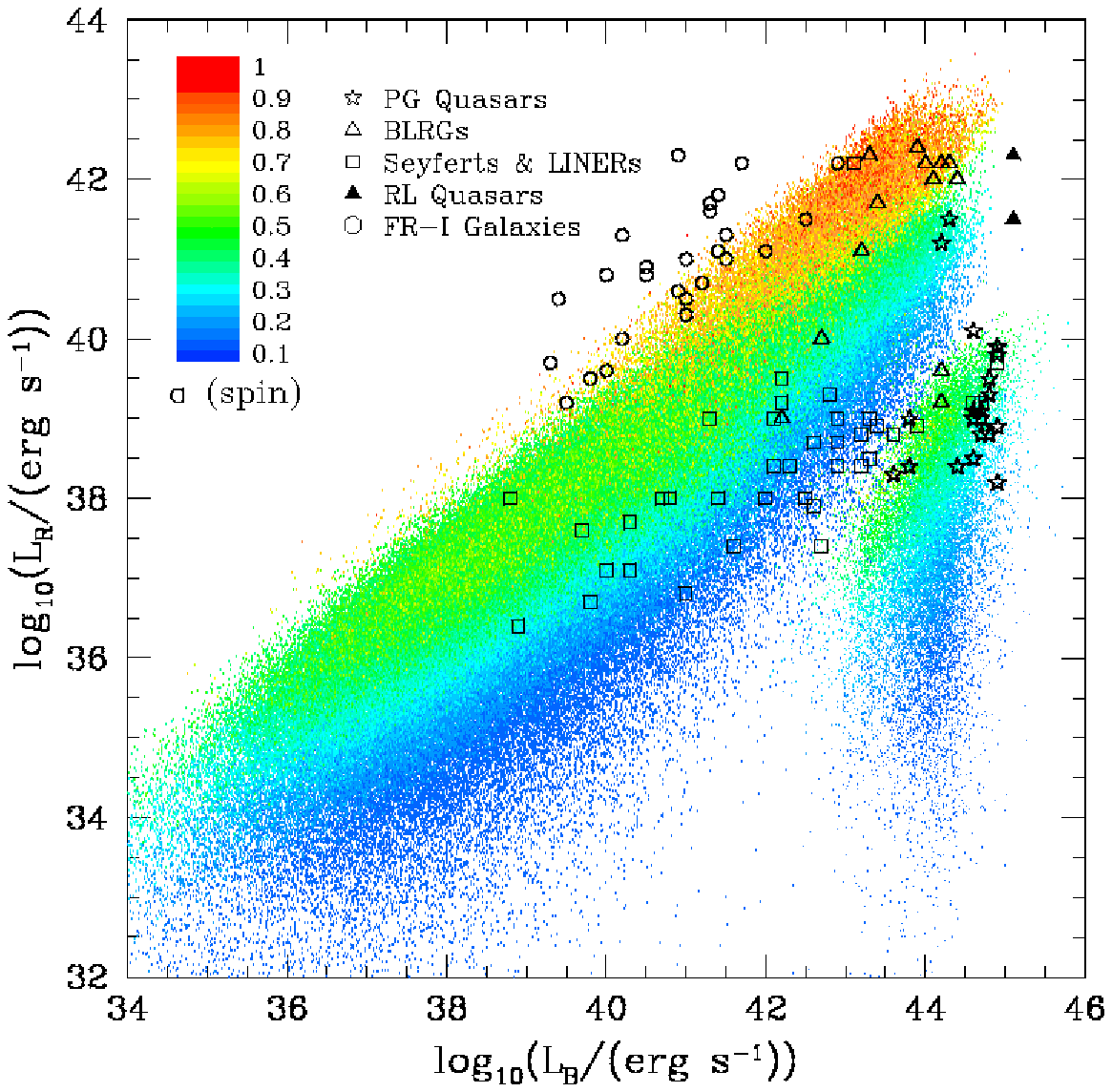}
\includegraphics[scale=0.365]{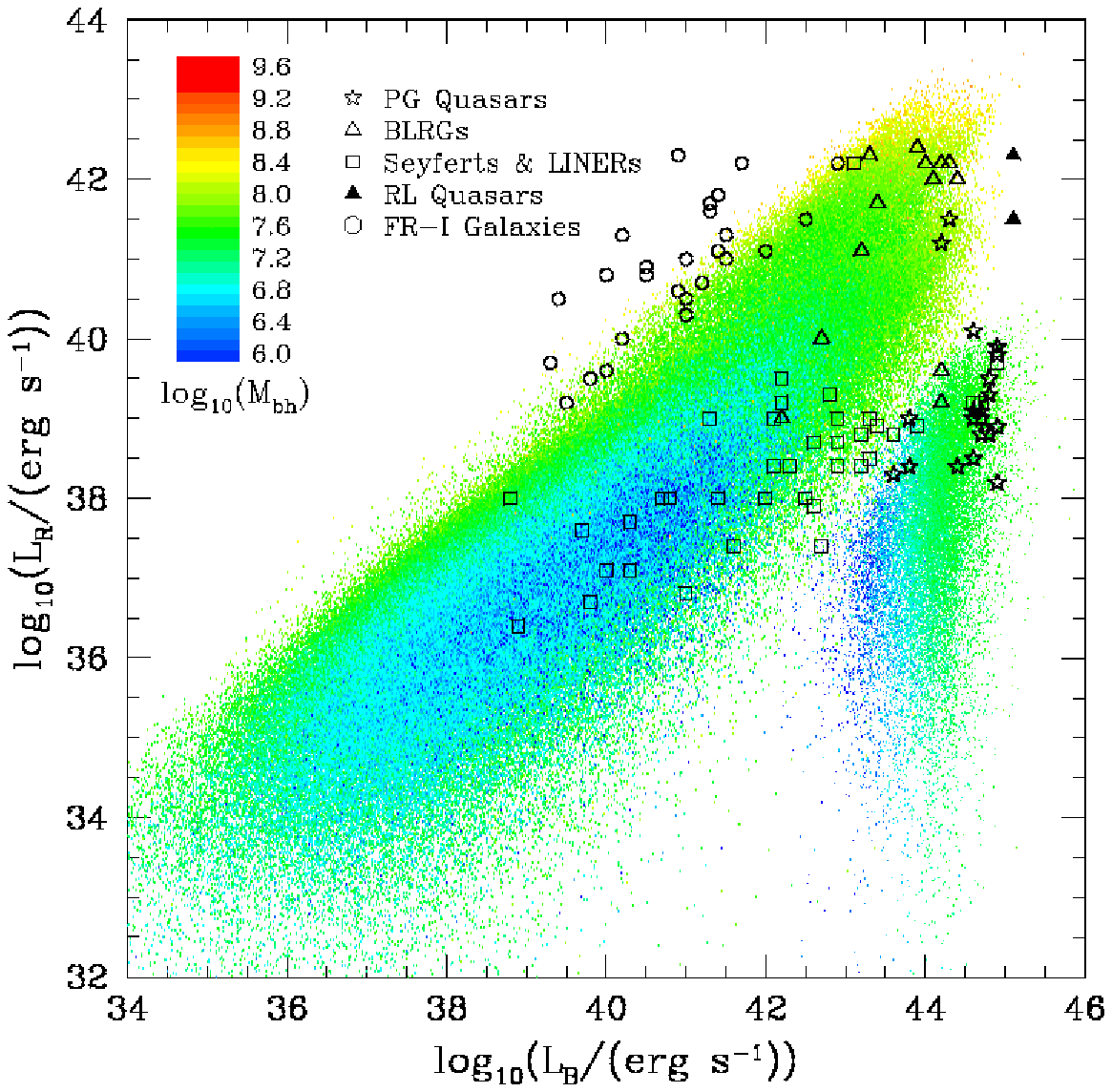} 
\includegraphics[scale=0.365]{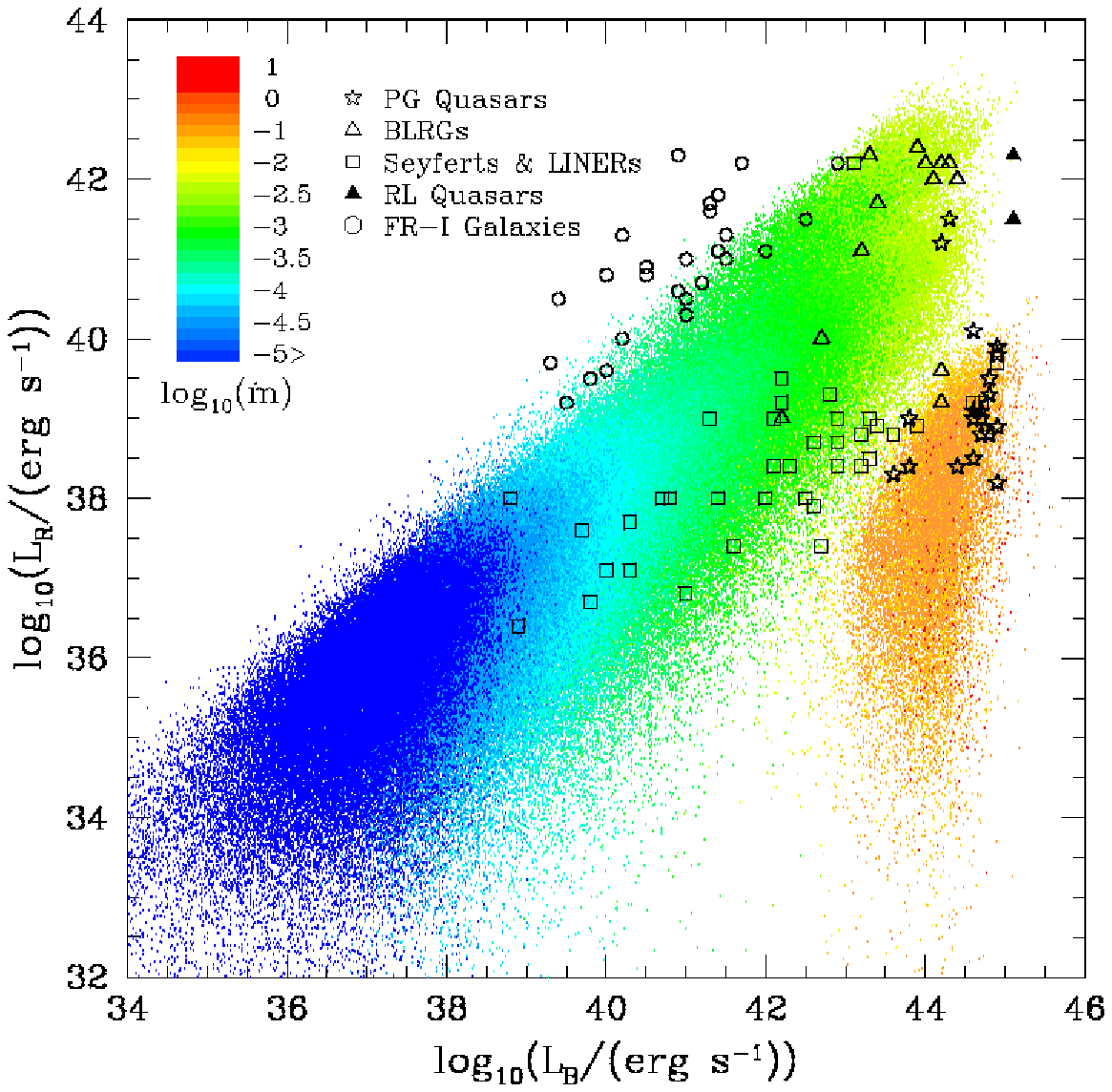}
\caption{Scatter plot of radio luminosity \textit{vs}. $B$-band
nuclear luminosity for the chaotic accretion model (as in
Fig.~\ref{jet_luminosities}), with model points coloured-coded
according to BH spin (top), mass (middle) and accretion rate
(bottom). The colours represent different ranges of values as
indicated in each key.}
\label{bz_parameters}  
\end{figure}

In order to gain further insight into the predictions of the model, we
show in Fig.~\ref{bz_parameters} how the main parameters that
determine the radio loudness of an AGN, $a$, $\MBH$ and $\dot{m}$, are
distributed across the $\log L_{\mathrm{B}}-\log L_{\mathrm{R}}$ plane
for the chaotic accretion model.

Comparing these plots to the data in Fig.~\ref{jet_luminosities},
the brightest radio-loud galaxies i.e. FR-I and BLRG, are identified
with very massive ($10^{8}-10^{9}~\Msun$), rapidly rotating BHs
(spins $\gtrsim0.8$) accreting at the top of the ADAF branch at
$\dot{m}\sim 0.01$. In our galaxy formation
model, these objects are hosted exclusively by giant ellipticals and their central BHs grow
through mergers and through the radio mode. The radio loud quasars (only
2 in our volume) are the extreme end of this population, and can be
matched by the highest mass objects in our sample. One issue with
identifying these objects with the ADAF branch is that they
typically have high excitation spectra, showing that there is a
bright UV disc (see e.g. Marchesini, Celotti \& Ferrarese
2004). However, the top of the ADAF branch is where the transition
to a thin disc takes place. Observations of stellar mass BH binary
systems show that this transition is complex, probably taking on a
composite structure with the thin disc replacing the hot flow at
progressively smaller radii (see e.g. the review by Done, Gierlinski
\& Kubota 2007). Thus, better modelling of the details of the
transition may well be able to reproduce the required UV ionising
spectrum.

The Palomar-Green (PG) Quasars are mostly concentrated at the top end of the narrow
thin disc sequence, extending from $\sim10^{43}$ to
$\sim10^{45}~\ergsec$ in optical luminosity and from $\sim10^{38}$ to
$\sim10^{40}~\ergsec$ in radio luminosity. The higher mass accretion
rates required to accrete via a thin disc are only generally possible
in lower mass BHs, $10^{7}-10^{8}~\Msun$ (see Fig.~\ref{mbh_mdot}), so these
have similar optical luminosities to the top of the ADAF branch as
their higher mass accretion rate is partially cancelled by the lower
mass BHs.  These lower mass BHs are hosted in lower mass galaxies,
predominantly spirals, and have lower spin, but the majority of the
suppression in radio power comes from the $\sim\negthinspace2$ orders
of magnitude stronger jet emission in the ADAF regime than in the thin
disc regime at $\dot{m}=0.01$. However, a few of the optically brightest
radio quiet objects are powered by Eddington (or even super-Eddington)
accretion onto $10^8~\Msun$ BHs, and these could be hosted by ellipticals.

Seyfert and LINER galaxies constitute an interesting sample in our model
because of their heterogeneity. Those exhibiting high nuclear
luminosities ($L_{\mathrm{B}}\gtrsim10^{43}~\ergsec$) (mainly
Seyferts), appear to occupy both the ADAF and thin-disc regimes,
whereas those with lower luminosities (LINERs) are generally powered
by an ADAF. These are the most numerous type of AGN seen in the local
Universe. The majority of these objects have BHs with masses of
$10^{7}-10^{8}~\Msun$. Thus, they have low-to-moderate spin and are relatively
quiet at radio luminosities. These are the same population of
objects  as the PG Quasars, but at lower mass accretion rates.

\section{Discussion}

The process of BH growth in our model is dominated by the accretion
of large amounts of gas that allow a BH to double its mass several
times during a Hubble time. The evolution of the spin of the BH is strongly
influenced by the nature of the accretion episodes. 
It is often assumed that BHs accrete gas via a
disc with constant angular momentum. In this case
(the prolonged accretion model), BHs are systematically spun up
during accretion, with most of them ending up with maximal
spin. Even minor mergers or disc instabilities, typical of the
growth of the bulge of spiral galaxies, can trigger a gas flow onto
the BH of mass comparable to that of the BH itself and thus lead to $a=1$. By contrast, if
the accreting material fragments at its self-gravity radius and the
associated star formation randomises the angular momentum direction
of each of the fragments (the chaotic accretion model), then
accretion proceeds in multiple {\em randomly oriented}
episodes. This results in a low spin BH. However, for the most
massive BHs, the major growth channel is not accretion but BH-BH
mergers, and these generate fairly rapid spins of $0.7-0.8$. Thus,
in the prolonged accretion model, the most massive BHs have slightly
lower spin than the bulk of the BH population, whereas in the
chaotic model they have larger spin.

Our results for the spin distributions are in good agreement with those of Berti \& Volonteri (2008). However, our predictions for the prolonged-accretion model deviate significantly from the results of Lagos \etal (2009). These authors obtain a strong spin bimodality when they consider accretion of gas via an accretion disc of constant angular momentum. The difference between the two approaches seems from the different amounts of cold gas available for accretion in the two models (see Kim \etal 2010 for a recent study of the cold-gas abundance in the Bower \etal 2006 model). In the original Bower \etal (2006) model and in the updated version that we have adopted in this study, the typical amount of cold gas accreted by a $10^{6}-10^{8}~\Msun$ BH is enough to spin it up to the maximum value. This is not true, however, for the Lagos \etal model where cold gas accretion includes several episodes with $M_{\mathrm{gas}}<\MBH$ that do not spin up the BH efficiently. The difference in the final spin distributions illustrates the impact of the underlying galaxy formation model on the inferred properties of the BHs at the centre of galaxies.

We have used the resulting distributions of BH mass, spin and mass
accretion rate to calculate the properties of AGN, assuming that any
accretion at a rate $\dot{m}<0.01$ proceeds via an ADAF, while
accretion at higher rates forms a standard thin disc.  The optical
luminosity is then fairly straightforward to calculate, but the radio
luminosity depends on the jet model. This is poorly understood, but
both observations and theoretical models agree that the collapse of
the thick ADAF into a thin disc leads to a similar collapse of the jet
emission by several orders of magnitude. Thus, there is already a
clear accretion mode switch in radio-loudness predicted by these
models (Maccarone, Gallo \& Fender 2003; Jester 2005).

The main issue then is how jet power couples to spin. In the
prolonged accretion model, there is a very small range in spin since all
BHs are spinning rapidly. Thus, {\em independently of the detailed jet
model}, the slope of the radio luminosity function is directly given
by the relative numbers of low and high mass BHs accreting in the ADAF
regime. This is steeper than observed since lower mass BH are much
more numerous than higher mass ones. Instead, in the chaotic accretion model, the observed radio luminosity
function can be reproduced {\em if the jet power depends strongly on
spin}. The $a^2$ dependence on spin power in the BZ jet
models is sufficient to make a large difference in radio luminosity
between the numerous low mass, low spin BHs and the much rarer high
mass, high spin BHs. This model reproduces the shape of the observed
radio luminosity function very well. 

A mass-spin correlation was also suggested by Sikora \etal (2007)
to explain the range in the radio-to-accretion disc luminosity ratio
seen in their (very heterogeneous) sample of AGN.  However, the model
they propose to establish this correlation is rather different from
ours which is based on an actual calculation of galaxy formation from CDM initial
conditions. Instead, Sikora \etal speculated that the high spin
resulted from major mergers which triggered large gas flows with
constant angular momentum direction onto the nucleus, spinning up the
BH and producing an elliptical galaxy. Conversely, they argued that
low spin resulted from minor mergers of randomly aligned satellite
galaxies (a random walk with spin up, spin down). These leave the
galactic gas disc intact, producing a spiral galaxy. However, as
discussed above, since most of the mass comes from the disc of the host galaxy, even minor mergers result in enough gas flowing to
the centre to spin the BHs up to maximal. Fragmentation and chaotic
accretion are necessary in order to produce low-spin BHs.  Major
mergers are indeed the key to the high spin of the most massive BHs,
but this is the result of BH-BH mergers, not of gas accretion.

One of the distinguishing features of our work is that the properties
of the BHs and their associated AGN are calculated {\it ab initio}
within the context of a full model of galaxy formation in a
$\Lambda$CDM universe. This model has been shown to agree with a large
variety of observational data such as galaxy luminosity functions in
various passbands and at different epochs, galaxy colours, the cosmic
star formation history and structural scaling properties such as the
Tully-Fisher relation and the $M_{\rm BH}-\sigma$ relations (Bower et
al. 2006). In this work we have augmented the galaxy formation model
with calculations of BH spin and of the optical and radio output
during accretion onto the BH. The two different accretion models we
have explored (prolonged and chaotic) span the range of likely spin
distributions and are good templates to investigate how the optical
and radio luminosities of accreting systems depend on BH spin. As
shown in Sections 5 and 6 and discussed in the previous paragraphs,
the chaotic model gives a reasonable overall match to the
observations, suggesting that a model in which BH spins develop a
bimodal distribution is the most plausible.

\section{A unification scheme for the AGN activity}

\begin{figure*}
\center
\includegraphics[scale=0.43]{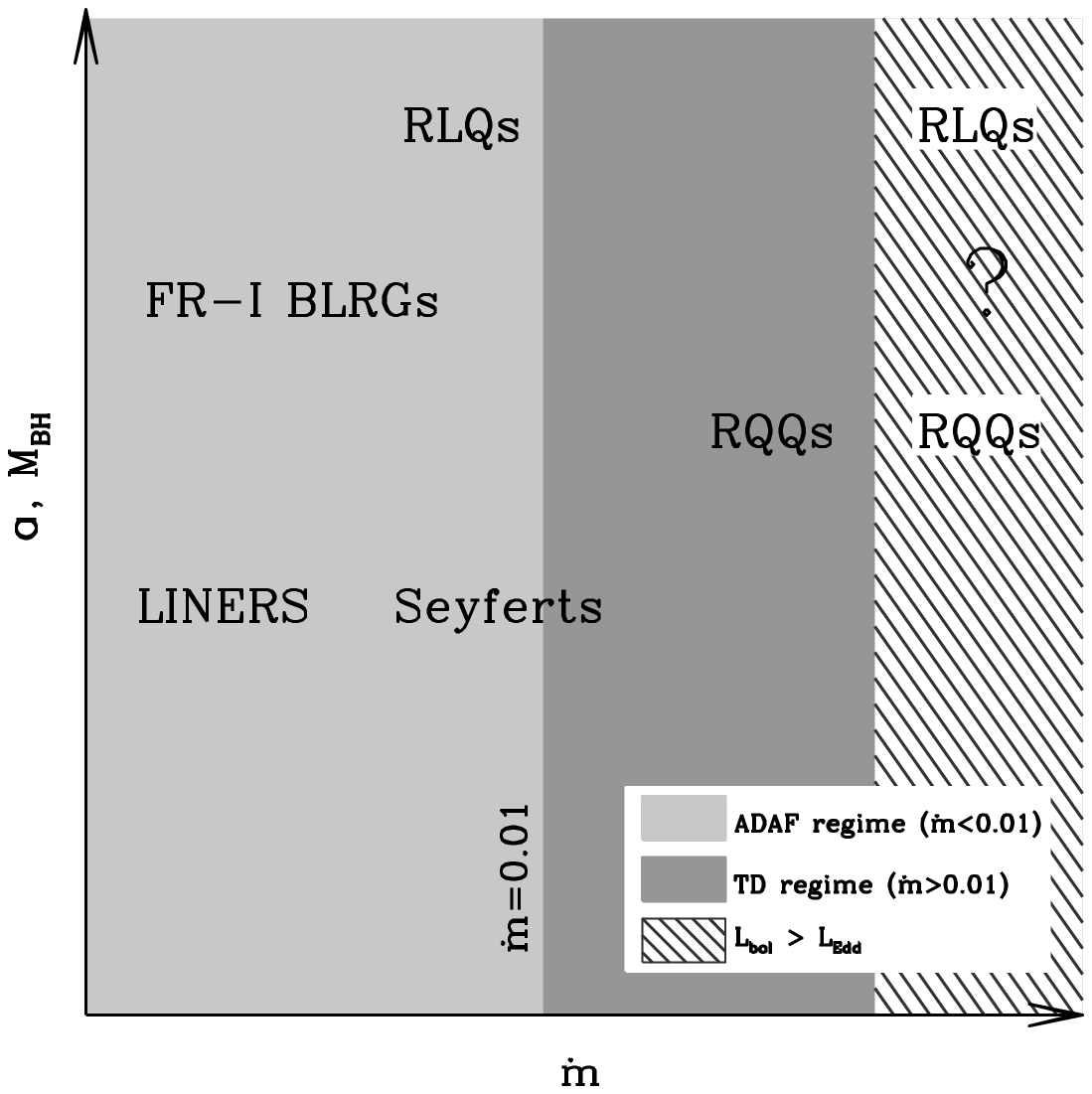}
\includegraphics[scale=0.43]{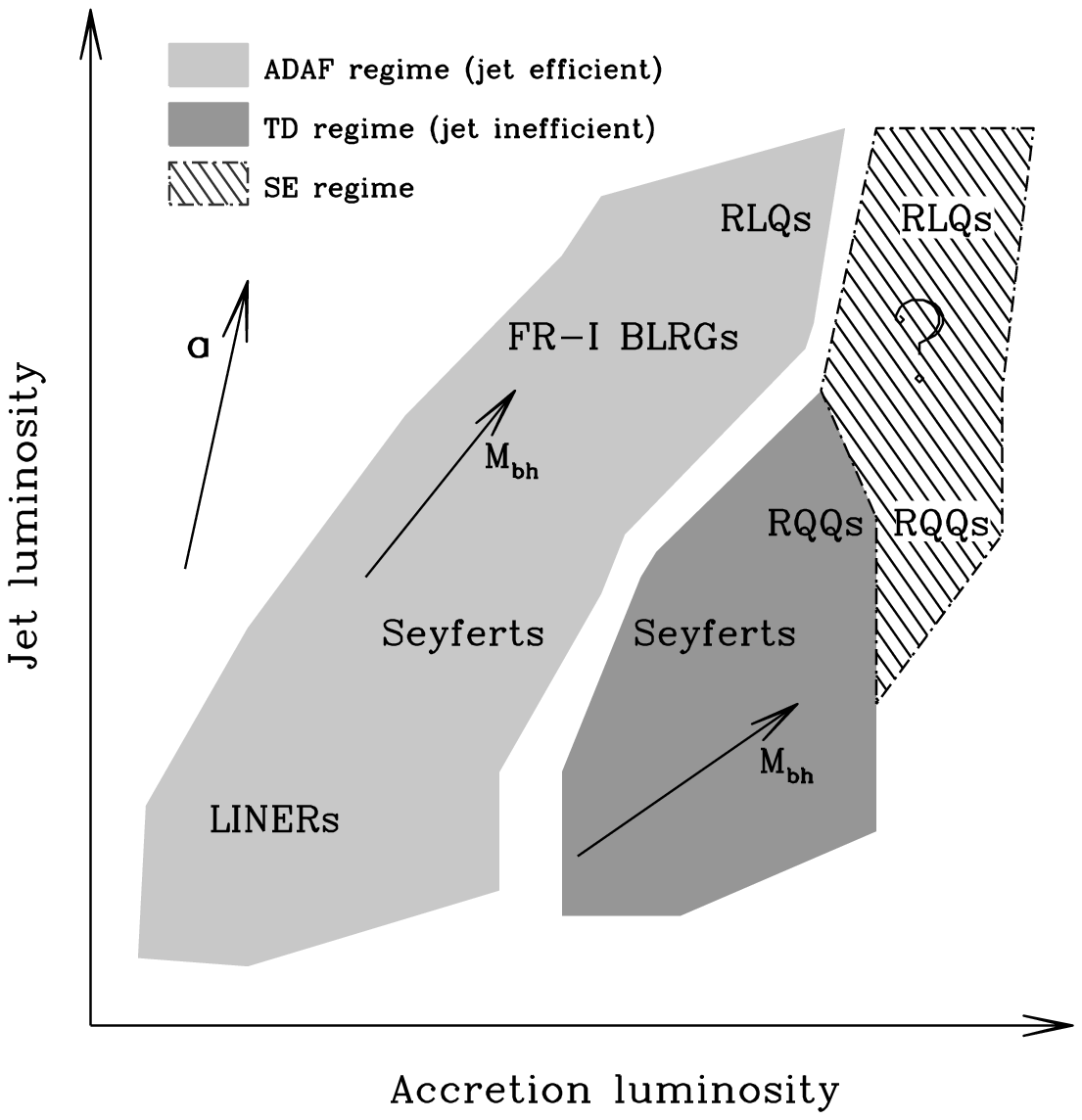}
\caption{Left: the location of the different AGN types on the
fundamental parameter ($a,\dot{m},\MBH$) plane. The $a$ and $\MBH$ 
axes have been merged into one since the two parameters are correlated 
as shown in the right hand plot of Fig.~\ref{mbh_spin} in the chaotic accretion model. 
The shaded areas
represent the different accretion regimes indicated by the keys. Right: the location of the
different AGN types on the optical -- radio plane. The shaded areas
represent different accretion regimes. The arrows show how the
physical parameters vary on the plane. Radio galaxies and LINERs lie in  the ADAF regime. 
Seyferts are preferentially found in the thin-disc regime,
even though a substantial population of Seyferts may be powered by
an ADAF. RQQs accrete at relatively higher accretion rates than
Seyferts and RLQs lie at the top of the ADAF branch. We speculate
that some of the latter may be found also in the super-Eddington
regime.} 
\label{unification}
\end{figure*}

Finally, we present an interpretation of the optical and radio
signatures of an active nucleus at different evolutionary stages of
the host galaxy within a chaotic accretion framework. This is illustrated
schematically in Fig.~\ref{unification}, where we show the relative
position of the different types of active galaxies on the
fundamental parameter ($a$, $\dot{m}$ and $\MBH$) plane (left panel; 
the $a$ and $\MBH$ axes have been merged into one since the two 
quantities correlate as shown in the right-hand panel of Fig.~\ref{mbh_spin}) and
the jet/accretion-luminosity plane (right panel). We also speculate on
the outcome of super-Eddington accretion. Such objects are rare in the
local Universe, but are increasingly important at higher redshift (see
Fig.~\ref{mdot}).

We start by considering the sources with the lowest mass BHs. During
a minor galaxy merger, cold gas from the disc is transferred to the
bulge, triggering star formation and BH growth. This process 
represents a natural mechanism for the gradual growth of spheroids in
spiral galaxies (Parry \etal 2009). Accretion of gas onto the BH usually occurs at
sub-Eddington rates through a thin disc which turns the host spheroid
into the bright nucleus of a Seyfert galaxy. The bright UV flux from
this disc produces a strongly ionising spectrum, resulting in strong
emission lines.  However, gas flows close to the disc may also
contribute to the emission (or absorption) spectrum of the
nucleus. The orientation of the central engine relative to a distant
observer may have an important impact on the observed spectral
features of the source and could account for the Type 1 and 2
sub-classification of Seyferts.

If the host galaxy experiences a major merger or a galactic disc
instability then the entire galactic gas disc is assumed to lose most
of its angular momentum, participate in a starburst and is added to 
the stellar spheroid mass. This supplies the central region of the galaxy with large
amounts of cold gas that feed the SMBH with several solar masses of
material per year at a near- or super-Eddington rate. Thus, the
nucleus becomes exceptionally luminous, resulting in a quasar. The
collapse of the cold gas reservoir also initiates an intense starburst which could
be contemporaneous with the quasar phase. However, during the quasar
phase, the galaxy is normally too dim to be seen against the vast
amounts of radiation produced by the central engine. The radio
output of the central engine when the accretion rate becomes
super-Eddington is not clear since our models do not extend to that
regime. Perhaps if the magnetic field strength close to the BH is 
enhanced due to a transition to a thick disc when the flow exceeds
the Eddington limit, strong jets may be launched establishing the
galaxy as a RLQ.

The duration of quasar activity is usually a few hundred million
years. At the end of it (and the associated starburst), the system
will have consumed or ejected its cold gas leaving a stellar
bulge. In most halos, the ejected gas falls back into the galaxy and
a new galactic disc is formed on a dynamical timescale. Further
gas accretion, or galaxy mergers, may subsequently trigger another
period of quasar activity and initiate a new growth era for the
SMBH.  Eventually, the halo may grow massive enough that gas is no
longer able to cool rapidly and the halo enters the hydrostatic
cooling regime. In this case, the end-product may be an elliptical
galaxy in which further gas cooling is restricted by the radio-mode
feedback. By construction, this accretion onto the BH takes place in
the ADAF regime and the host elliptical appears as a radio
galaxy. In the extreme case where the accretion occurs at the top of
the ADAF branch and the central SMBH is rapidly rotating the
galaxy is identified as a RLQ.  The orientation of the central
engine and jets may explain the different subcategories of RLQs. For
example, if the jet axis lies close to the line-of-sight of a
distant observer, the source may be visible as a blazar.

\section{Conclusions}

We have presented a model of AGN activity in which the evolution of
the galaxy is calculated in its full cosmological context and the
luminosity of AGN at radio and optical wavelengths is calculated
using a model for gas accretion and the generation of jets. We first
considered the evolution of SMBH spin. We have found that the
different astrophysical processes that influence the growth of SMBHs
have a significant effect on the global spin distribution. For
example, if accretion of gas of constant angular momentum dominates
the growth of SMBHs, the associated holes will be rapidly rotating. 
However, if the gas that is fed into the SMBH has random angular
momentum, a bimodal spin distribution results. In this case, high spin values occur only
for the most massive BHs ($\gtrsim10^8~\Msun$), and these are mainly
due to gas poor major mergers, where the BH growth is dominated by
the BH-BH merger.

We have coupled this mass, spin and mass accretion rate evolution to a
model for the accretion flow and jet. The accretion flow is assumed
to form a geometrically thick, hot, radiatively inefficient flow
(ADAF) for $\dot{m}<0.01$, and to collapse to a thin disc at higher
mass accretion rates. The jet power couples strongly to the
accretion mode since it depends on the vertical (poloidal) magnetic
field component close to the BH horizon. The collapse by two orders
of magnitude in the scale height of the flow results in a similar
drop in radio power. This already produces a dichotomy in radio
properties which explains the distinction between radio-loud and radio
quiet objects.

However, we also find that for our model to match the slope of the
observed radio luminosity function, it is necessary for low mass BHs
to generate relatively less jet power than high mass BHs. This is
readily achieved in models where the jet couples {\em strongly} to
spin, such as in the classic BZ jet mechanism, {\em
and} where lower mass BHs have lower spin than the most massive BHs,
as in our chaotic accretion model.

Coupling the chaotic accretion model to the BZ jet mechanism results
in an AGN population which reproduces the diversity of nuclear
activity seen in the local Universe. In particular, the model
accounts for the radio and optical luminosities of the FR-I, BLRG,
Seyfert and LINER galaxy populations. This is the first consistent 
demonstration that a great part of the phenomenology of AGN can be
naturally explained by the coeval evolution of galaxies and BHs,
coupled by AGN feedback, in a cold dark matter universe.  In future
work, we will address the evolution of AGN across cosmic time, as a
crucial test of galaxy formation models.

\section*{Acknowledgements}
We thank an anonymous referee for useful suggestions that have improved our paper. NF acknowledges receipt of a fellowship funded by the European
Commission's Framework Programme 6, through the Marie Curie Early Stage
Training project MEST-CT-2005-021074. AJB acknowledges the support of the Gordon 
\& Betty Moore Foundation. CSF acknowledges a Royal Society Wolfson
Research Merit Award. This work was supported in part by an STFC
Rolling Grant to the Institute for Computational Cosmology.

\end{document}